\documentclass[lettersize,journal]{IEEEtran}
\usepackage{amsmath,amsfonts,amssymb}
\usepackage{algorithmic}
\usepackage{array}
\usepackage{cite}
\usepackage[caption=false,font=normalsize,labelfont=sf,textfont=sf]{subfig}
\usepackage{textcomp}
\usepackage{stfloats}
\usepackage{url}
\usepackage{verbatim}
\usepackage{graphicx}
\def\BibTeX{{\rm B\kern-.05em{\sc i\kern-.025em b}\kern-.08em
    T\kern-.1667em\lower.7ex\hbox{E}\kern-.125emX}}
\usepackage{balance}

\begin{document}
\bstctlcite{IEEEexample:BSTcontrol}
\title{Algorithms for Non-Negative Matrix Factorization on Noisy Data With Negative Values}
\author{Dylan Green and Stephen Bailey
\thanks{Original version submitted November 2, 2023; final version accepted September 23, 2024. The work of Dylan Green was supported by the U.S. Department of Energy (DOE), Office of Science, Office of Workforce Development for Teachers and Scientists, Office of Science Graduate Student Research (SCGSR) program. The SCGSR program is administered by the Oak Ridge Institute for Science and Education for the DOE under contract number DE‐SC0014664.
Stephen Bailey is supported by the U.S. DOE, Office of Science, Office of High-Energy Physics, under Contract No. DE–AC02–05CH11231.  This work used the National Energy Research Scientific Computing Center (NERSC), a DOE Office of Science User Facility under the same contract. \textit{(Corresponding author: Dylan Green)}

Dylan Green is with the Department of Physics and Astronomy, University of California Irvine, Irvine, CA 92697 (email: dylanag@uci.edu)

Stephen Bailey is with Lawrence Berkeley National Laboratory, Berkeley, CA 94720 (email: StephenBailey@lbl.gov)
}}

\markboth{}{2}
\markboth{Draft version September 20, 2024}{}

\maketitle

\begin{abstract}
Non-negative matrix factorization (NMF) is a dimensionality reduction technique that has shown promise for analyzing noisy data, especially astronomical data.
For these datasets, the observed data may contain negative values due to noise even when the true underlying physical signal
is strictly positive.  Prior NMF work has not treated negative data in a statistically consistent manner, which becomes
problematic for low signal-to-noise data with many negative values.
In this paper we present two algorithms, Shift-NMF and Nearly-NMF, that can handle both the noisiness of the input data and also any introduced negativity. 
Both of these algorithms use the negative data space without clipping or masking and recover non-negative signals without any introduced positive offset that occurs when clipping or masking negative data. 
We demonstrate this numerically on both simple and more realistic examples, and prove that both algorithms have monotonically decreasing update rules. 
\end{abstract}


\section{Introduction}
\IEEEPARstart{N}{on-negative} matrix factorization (NMF) \cite{paatero_positive_1994} is a powerful matrix factorization method
that has gained significant traction as a dimensionality reduction technique. The NMF specification
is, broadly speaking, to fit coefficients $\mathbf{H}$ and templates $\mathbf{W}$ to the (possibly noisy) data $\mathbf{X}$ such that 
\begin{equation} \label{eq:vanilla}
    \mathbf{X} \approx \mathbf{W}\mathbf{H},
\end{equation}
with the constraints $\mathbf{X}, \mathbf{W}, \mathbf{H} \in \mathbb{R}_{+}$, i.e. all values in 
all matrices are non-negative.
In this formalism, $\mathbf{X}$ is a $d \times n$ matrix of $d$-dimensional data with each of the $n$ 
individual data vectors on the columns of $\mathbf{X}$. $\mathbf{W}$ is therefore $d \times q$ 
and $\mathbf{H}$ is $q \times n$ where $q$ is the desired number of templates to fit. 
Approximating $\mathbf{X}$ necessitates defining a measure of closeness between the
reconstruction and the original dataset. Here we are concerned with the Euclidean norm, which
allows us to cast the problem as minimizing the objective function
\begin{equation} \label{eq:objective}
    ||\mathbf{X} - \mathbf{WH}||^2
\end{equation}
with the aforementioned constraints. A common method for solving this matrix optimization
problem is to use multiplicative iterative update rules, such as those originally derived by
Lee and Seung \cite{lee_algorithms_2000}.

The applications for NMF as a data reduction technique are broad, and cover a variety of fields, for example geophysics  \cite{pauca_nonnegative_2006}, biology \cite{li_non-negative_2013}, and written character processing \cite{lin_novel_2019} to name a few. Astrophysics in particular has recently
started to explore new and exciting applications of the method and its possible implications
for data processing. In astronomical imaging and spectroscopy, the signal features
are strictly non-negative and the constraints of the method become 
attractive. However, even though the \textit{true} signal will be strictly non-negative,
the \textit{observed} data may be negative due to noise.
This noise could arise from instrumental readout,
from noisy calibrations, or from the subtraction of a mean value of a noisy background.
Regardless of the source of the noise, it is important to note that downward fluctuations of
the signal are just as valid as upward fluctuations, even when those downward fluctuations
result in negative data.  Any analysis of the data should treat both positive and negative values in a statistically consistent manner.

Even if all data values are positive, the noise values intrinsic in data collection mean that standard NMF as presented
in (\ref{eq:vanilla}) will perform suboptimally and will attempt to fit noise
values when generating template and coefficient matrices. Prior work to handle these noise values
has generally been focused on adequately weighting the fitted templates in an attempt to down
weight the noisiest data, and up weight the data closest to noise-free, without concern for the negativity. This has been historically
successful, with the first development along these lines being the specific
NMF variant  derived in \cite{blanton_k_2007}, which uses NMF to model galaxy spectra. In this methodology, the NMF fit is weighted and the trained template
vectors are constrained to be a non-negative combination of a larger set of non-negative
and smooth input templates. Generally these weights are set to non-negative values such that
locations where the data is negative get less or zero weight compared to the positive values
of the data matrix. Both the weighting of the data pixels as well as the smooth and 
non-negative input
templates serve to prevent the algorithm from overfitting the noise values of the observed data.
This is ultimately a weighted version of non-negative matrix tri-factorization, a methodology
similar to NMF
that factorizes $\mathbf{X}$ into three matrices (rather two) such that
\begin{equation} \label{eq:trifactor}
    \mathbf{X} \approx \mathbf{GWH}.
\end{equation}
In \cite{blanton_k_2007}, $\mathbf{G}$ is initialized and held fixed through the updates. For more details on 
non-negative matrix tri-factorization see \cite{copar_fast_2019}. 

Other similar developments include the Heteroskedastic Matrix Factorization method in \cite{tsalmantza_data-driven_2012}, which does not restrict the basis vectors to any
subspace but does require them to be non-negative and weights the data. 
While not directly identified as a variant of NMF, the update rules
presented match exactly those of an NMF variant. A vectorized version of the same update
rules was independently derived in \cite{zhu_nonnegative_2016}. Both of these methods
were explicitly also tested on 1-dimensional galaxy spectra. The update rules presented in \cite{zhu_nonnegative_2016}
were used for 2-dimensional images of circumstellar disks in \cite{ren_non-negative_2018} and were
extended for use on a larger dataset through the use of GPUs in \cite{m_nmf-based_2023}. More detail on this method, which is the most direct precursor to the algorithms presented in this paper, is presented in Section~\ref{sec:weight-nmf}.

An alternate and ultimately quite different method to handle noisy data is proposed in \cite{boulais_unmixing_2021},
where noisiness is handled by repeating a regularized but unweighted NMF fit to the noisy data
many times and using a Monte
Carlo analysis of these fits to determine the optimal final basis set of templates. Corresponding coefficients
for the final templates are then found using a non-negative least squares solver. 
This too was also tested on 1-dimensional galaxy spectra. 

Weighting the data pixels to account for noise is not unique to the astronomy field. 
\cite{wang_ls-nmf_2006} derived weighted update rules for NMF for 
analyzing gene expression data by reformulating the
problem using a least squares approach. Subsequently in \cite{plis_correlated_2009} a
version of NMF is derived where the noise covariance is not
diagonal (as has been assumed in the preceding papers). 

One notable issue with all stated prior work in this field, however, 
is that standard NMF, even when weighted, does not admit \textbf{any} negativity in its data matrix $\mathbf{X}$. 
Even though weighting has shown promise in handling the noisiness of the input data, it 
does not explicitly handle the negative values that can be introduced by the noise. 
When using update rules, negative values need to be handled before fitting any templates and 
coefficients or else the negative values will pollute the updates and break the non-negativity
constraints on $\mathbf{H}$ and $\mathbf{W}$.
Common methods to handle negative data include clipping all negative values to 0 or to set
the weight of those values to 0, as is done in \cite{tsalmantza_data-driven_2012, zhu_nonnegative_2016, blanton_k_2007};
removing negative data from the input data matrix entirely as in \cite{ren_non-negative_2018, m_nmf-based_2023};
or to clipping the negative values to some small positive value $\epsilon \lesssim 10^{-6}$  as in \cite{boulais_unmixing_2021}. The support for negativity in the input data in public NMF code is varied. 
Scikit-Learn \cite{noauthor_sklearndecompositionnmf_nodate} does not admit any negativity in its input matrix, and throws an exception if any is detected. 
NonnegMFPy \cite{zhu_nonnegative_2016} clips any negative data in the input matrix to zero. R's 
built in NMF package \cite{noauthor_nneg_nodate} has the option to clip negative data, shift
the entirety of the matrix to be nonnegative, take the absolute value of the input data, or 
fit the positive and negative elements separately. All of these codes and methods do not handle
the negative values of the input matrix consistently with the positive values.

In high signal regimes where the signal dominates over all noise
incorrect handlings of negative data might have only a small and perhaps unnoticeable effect. 
However, when we push towards low signal to noise ratio (SNR) 
data where the noise-free signal is extremely close to 0 and the noisiness might
introduce negativity 
the weakness of this handling becomes evident. 

A simple test elucidating the weakness of this handling is to simulate data with an expected
positive signal near 0, where the noise pollutes the data with negative values.
To test we generate 500 unique instances, henceforth referred to as exposures, of an unnormalized double Gaussian feature. The true noise-free flux is represented by a function of pixel $x$,
\begin{equation} \label{eq:double_gauss}
    f_{\text{true}}(x) = A \exp\left({-\frac{(x - \mu_1)^2}{2 \sigma^2}}\right) + B \exp\left({-\frac{(x - \mu_2) ^2}{2 \sigma^2} }\right),
\end{equation}
where the two Gaussians have the same variance $\sigma^2$ 
but with different means $\mu_1$  and $\mu_2$. The
ratio of the amplitudes of the Gaussian features is controlled by the scaling constants $A$
and $B$, and varies between 1:2 and 2:1, with the ratio
randomly chosen for each exposure. We use the noise-free true signal from (\ref{eq:double_gauss}) as the mean of a Poisson
random number in each pixel to introduce Poisson noise before then adding zero-mean Gaussian noise to generate our noisy exposure $\tilde{f}$:
\begin{align} \label{eq:simple_test}
\begin{split}
    \tilde{f}(x) &= P + N\\
    P &\sim \text{Pois}( f_{\text{true}}(x)) \\
    N &\sim \mathcal{N}(0, \sigma_{\text{noise}}^2)
\end{split}
\end{align}
The variance of the Gaussian noise, $\sigma_{\text{noise}}^2$,
is picked uniformly and randomly such that each exposure has a different variance.

We use the estimated inverse variance of each pixel as weights in the weighted NMF update
rules (\ref{eq:update_weight}) where we estimate the variance by adding the 
variance of the Gaussian noise and the estimated variance of the signal from the Poisson process:
\begin{equation}
    \tilde{\sigma}^2(\lambda) = P + \sigma_{\text{noise}}^2
\end{equation}

By design the noise will drive some pixels negative outside of the Gaussian doublet. 
To handle this we apply the scheme where we clip all negative values to 0, as is done
in the code released alongside \cite{zhu_nonnegative_2016}. 
We then generate 2 NMF templates from these 500 exposures, and attempt to reconstruct the
input from those two templates. 

\begin{figure}[!t]
\centering
\includegraphics[width=\columnwidth]{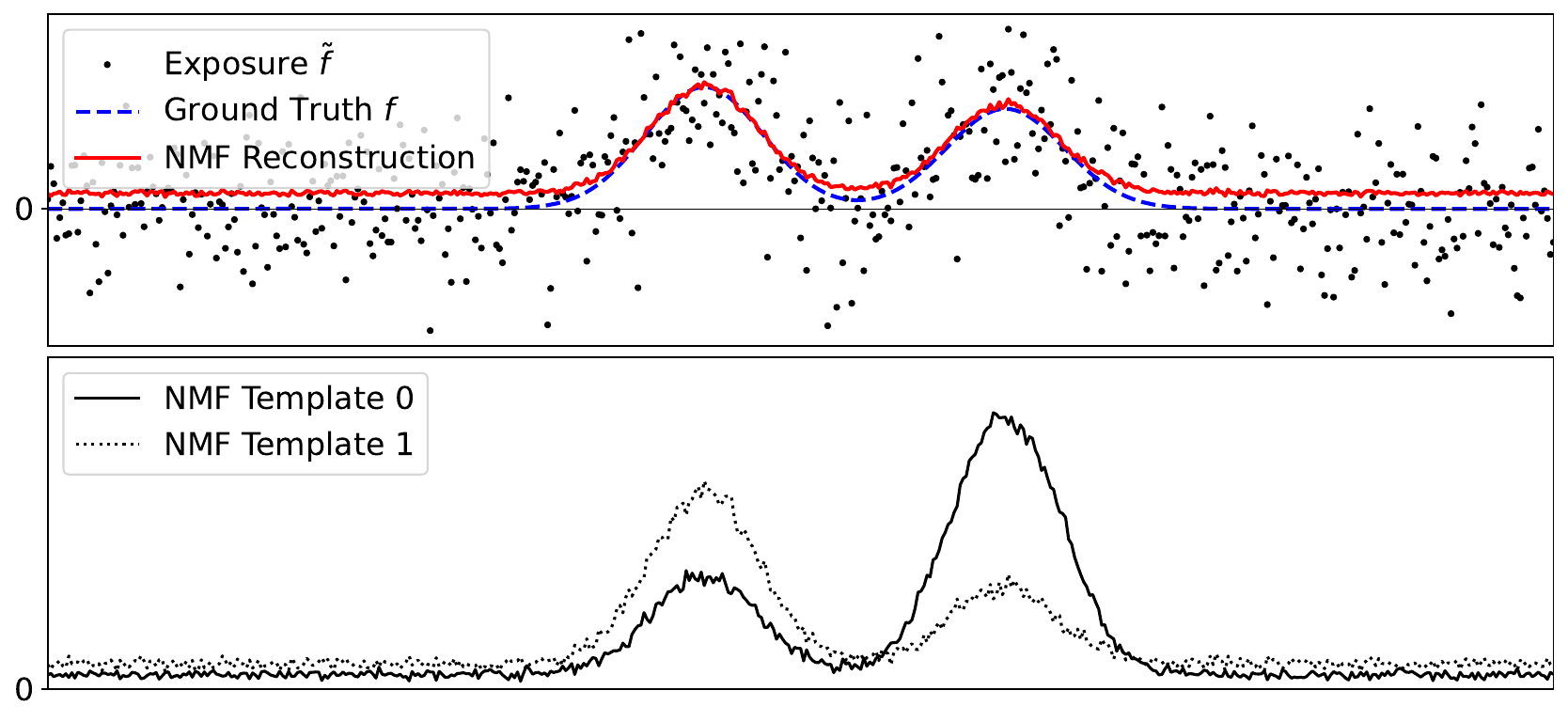}
\caption{Results of two weighted NMF templates generated on a toy example. See text for details of toy example. The top panel shows one representative exposure from the set of 500 as dots, with the noiseless truth and the template based reconstruction overplotted in dashed blue and solid red respectively. The bottom panel shows the two raw templates, one dotted and one solid, scaled so the maximum value is 1 but preserving the relative scale between the two templates. Notice that the templates, both the reconstruction in the upper panel and the templates themselves in the lower panel, have a positive vertical offset in the region where the truth is 0 due to the templates only fitting the positive component of the noisy data.}\label{toy_templates}
\end{figure}

Fig. \ref{toy_templates} shows the results of this simple test. The top panel shows a representative
exposure from the data set used to generate the templates in dots, 
as well as the noiseless truth used to generate it (in blue and dashed) and the template based reconstruction
(in red and solid). 
The bottom panel shows the two templates generated after 100 iterations of updates, one dotted and one solid. 
It is clear even in this extremely simple example that the templates introduce a positive offset from zero
outside of the doublet when there is no such baseline in the true data. Since the
NMF fit has only been allowed to fit the positive component of the noisy data, it has 
incorrectly estimated that a best fit must have a positive, non-zero, continuum in that region.  

\subsection{Our Contributions}
While the standard NMF update rules cannot correctly handle negative values of the data, it \textit{is} possible to
consider the negative components of the data in the fit while still maintaining the non-negativity
constraint on the coefficients and templates. Our contributions in this paper amount to two novel
algorithms with the capability to correctly account for negative data elements
caused by noise while still generating strictly 
non-negative coefficients and template matrices. We have named these two algorithms Shift-NMF and Nearly-NMF
based on their method of fitting. Both of these algorithms additionally
include weighting, which allows both algorithms to be more robust to the noise itself
and to handle missing data with weights equal to 0.
We will
derive update rules for both of these algorithms akin to those presented in \cite{lee_algorithms_2000}.

After defining the algorithms, we will demonstrate their effectiveness on the same toy problem
presented in this Introduction. After validating their results on a small scale test, we
will expand and validate the results with a more realistic simulated test motivated by applications to
astrophysics datasets. 
This simulated data is designed to emulate the scale and depth of the data that these
methods might realistically be used for. This test will also motivate some brief 
studies of the algorithms' computational properties. Our closing remarks will summarize these
results, while also presenting some possible extensions as well as potential applications
of these methods to fields outside of astrophysics.

\subsection{Notation}
Throughout this paper matrices will be represented by bold uppercase letters ($\mathbf{X, H, W}$), 
vectors by lowercase bolded greek letters ($\boldsymbol{\alpha, \beta}$) and scalars by lowercase unbolded letters ($a, b, \sigma$).

The notation $\mathbf{A}_{,j}$ will refer to the $j$-th column of  $\mathbf{A}$, which is a vector. The notation
$[\mathbf{A}]^+$ refers to a matrix the same shape as $\mathbf{A}$, where all negative values are set to zero
and the positive values maintained. Correspondingly $[\mathbf{A}]^-$ is a matrix containing
the absolute value of the negative elements of  $\mathbf{A}$, with the formerly positive values
set to 0, such that therefore $\mathbf{A} = [\mathbf{A}]^+ - [\mathbf{A}]^-$.

For all following sections the open circle dot operator $\circ$ is used for the
Hadamard (elementwise) product and the division bar is applied elementwise.

\section{Overview of Algorithms}
\subsection{Introduction to NMF with Heteroskedastic Weights}\label{sec:weight-nmf}

Lee and Seung, in \cite{lee_algorithms_2000}, derived multiplicative update rules for 
minimizing (\ref{eq:objective}), where each matrix $\mathbf{H}$ and $\mathbf{W}$ are
updated elementwise and independently of each other by multiplying each matrix elementwise
by a multiplicative factor:
\begin{align}
\begin{split}\label{eq:update_vanilla}
    \mathbf{H} &\leftarrow \mathbf{H} \circ \frac{\mathbf{W}^T \mathbf{X}}{\mathbf{W}^T \mathbf{WH}} \\
    \mathbf{W} &\leftarrow \mathbf{W} \circ \frac{\mathbf{X}\mathbf{H}^T}{\mathbf{WH}\mathbf{H}^T}
\end{split}
\end{align}
Given non-negative initializations for $\mathbf{W, H}$ these update rules guarantee that
the final matrices $\mathbf{W, H}$ will remain non-negative as the multiplicative
update factor to each matrix 
($\frac{\mathbf{W}^T \mathbf{X}}{\mathbf{W}^T \mathbf{WH}}$ for $\mathbf{H}$ and
$\frac{\mathbf{X}\mathbf{H}^T}{\mathbf{WH}\mathbf{H}^T}$ for $\mathbf{W}$) is itself guaranteed to be non-negative. For a perfect
reconstruction of the input matrix, $\mathbf{X} = \mathbf{WH}$, the multiplicative factor
is identically equal 1.

In \cite{zhu_nonnegative_2016} Zhu introduces a \textit{weighted} version of the objective (\ref{eq:objective}),
\begin{equation} \label{eq:objective_weight}
    \chi^2 = ||\mathbf{V}^{1/2} \circ (\mathbf{X} - \mathbf{WH}) ||^2,
\end{equation}
where the matrix $\mathbf{V}$ is a matrix of the same shape as $\mathbf{X}$ containing 
heteroskedastic weights for each matrix element, with the square root applied elementwise.
The constraints are the same as (\ref{eq:objective}), $\mathbf{X}, \mathbf{W}, \mathbf{H} \in \mathbb{R}_{+}$,
with the added constraint $\mathbf{V} \in \mathbb{R}_{+}$. A common choice of weights is to use the inverse variance of each matrix element in $\mathbf{X}$.
Zhu correspondingly derives
update rules analogous to (\ref{eq:update_vanilla}) with the weighting matrix included:
\begin{align}
\begin{split}\label{eq:update_weight}
    \mathbf{H} &\leftarrow \mathbf{H} \circ \frac{\mathbf{W}^T (\mathbf{V} \circ \mathbf{X})}{\mathbf{W}^T (\mathbf{V} \circ (\mathbf{WH}))} \\
    \mathbf{W} &\leftarrow \mathbf{W} \circ \frac{(\mathbf{V} \circ \mathbf{X})\mathbf{H}^T}{(\mathbf{V} \circ (\mathbf{WH}))\mathbf{H}^T}
\end{split}
\end{align}

\subsection{Algorithm 1: Shift-NMF}\label{sec:shift-nmf}

A straightforward extension to account for negative data is to use a ``shift and
deshift'' methodology, which allows template fitting to consider the 
influence of normally negative values when fitting templates by shifting
both the data \textbf{and} the templates to the positive half space while keeping the
shift parameter separate and fixed in the fit.
This is easily interpreted as a two step process:

\begin{enumerate}
\item
  Shift the data by a value $y \geq 0$ so that all elements are non-negative. 
  $y$ will depend on the input matrix, and should be equal to
  or larger than the absolute value of the most negative value in the data matrix being fit, such that the new minimum
  value of the dataset after the shift is $\geq 0$: $y = |\min(0, \min(\mathbf{X}))|$.
\item
  Fit the desired number of NMF templates to the shifted data with the additional fixed shift term included.
\end{enumerate}

To derive update rules we start by defining the objective function to minimize with heteroskedastic weights $\mathbf{V}$
included as 
\begin{equation} \label{eq:objective_shift}
    \chi^2_{\text{SHIFT}} = ||\mathbf{V}^{1/2} \circ ((\mathbf{X} + y) - (\mathbf{WH} + y)) ||^2
\end{equation}

Note that both steps above are represented in this objective function: the shifted data
in step 1 is represented by $(\mathbf{X} + y)$, and the NMF reconstruction with a fixed
shift term is represented by $(\mathbf{WH} + y)$. We want to underscore that the fixing of the 
shift in the fitting process represents a fundamentally different 
solution to the NMF problem than fitting only a set of
free templates to the shifted dataset. Fitting NMF templates to a dataset that has been
shifted by an arbitrary constant $y$ to be entirely nonnegative is
represented by the objective function
\begin{equation*} \label{eq:objective_shift_no_hold}
    \chi^2 = ||\mathbf{V}^{1/2} \circ ((\mathbf{X} + y) - \mathbf{WH}) ||^2,
\end{equation*}
producing a distinctly different minima than (\ref{eq:objective_shift}). 
The resultant factorization that minimizes this objective specifically
minimizes the reconstruction of the shifted data $(\mathbf{X} + y)$, 
whereas with some light algebra it 
can be shown that minimizing (\ref{eq:objective_shift}) is identical to minimizing
(\ref{eq:objective_weight}). Thus the factorization produced by Shift-NMF attempts
to reconstruct the original data $\mathbf{X}$, \textit{not} the shifted data $(\mathbf{X} + y)$. 

In order to derive update rules we will follow the methodology of \cite{lee_algorithms_2000} by taking the derivative of (\ref{eq:objective_shift})
with respect to each of the two matrices $\mathbf{W}$ and $\mathbf{H}$ to get
\begin{equation}
\begin{aligned}
    \frac{\partial \chi^2}{\partial \mathbf{H}} &= -2 \mathbf{W}^T(\mathbf{V}\circ (\mathbf{X} + y) - \mathbf{V}\circ(\mathbf{WH} + y)), \\
    \frac{\partial \chi^2}{\partial \mathbf{W}} &= -2 (\mathbf{V}\circ (\mathbf{X} + y) - \mathbf{V}\circ(\mathbf{WH} + y)) \mathbf{H}^T, \\
\end{aligned}
\end{equation}
and substituting them into update rules of the forms $\mathbf{H} \leftarrow \mathbf{H} - \mathbf{A}_{\mathbf{H}} \circ \frac{\partial \chi^2}{\partial \mathbf{H}}$, $\mathbf{W} \leftarrow \mathbf{W} - \mathbf{A}_{\mathbf{W}} \circ \frac{\partial \chi^2}{\partial \mathbf{W}}$, 
where $\mathbf{A}_{\mathbf{H}}$  and $\mathbf{A}_{\mathbf{W}}$ are learning rate hyperparameters. 
For constant $\mathbf{A}_{\mathbf{H}}$  and $\mathbf{A}_{\mathbf{W}}$ these update
rules correspond to standard gradient descent, but we can instead set adaptive learning
rates to make these multiplicative update rules. In this case, the adaptive learning
rates are given by
\begin{equation}
\begin{aligned}
    \mathbf{A}_{\mathbf{H}}  &= \frac{\mathbf{H}}{2\mathbf{W}^T(\mathbf{V}\circ(\mathbf{WH} + y))}\\
    \mathbf{A}_{\mathbf{W}}  &= \frac{\mathbf{W}}{2(\mathbf{V}\circ(\mathbf{WH} + y)\mathbf{H}^T)}
\end{aligned}
\end{equation}
which are adapted from the original adaptive learning rate for $\mathbf{H}$ given in
\cite{lee_algorithms_2000} to include the weight matrix $\mathbf{V}$ and the shift $y$.
With some rearranging, this substitution produces the final update rules
\begin{align}
\begin{split}\label{eq:update_shift}
    \mathbf{H} &\leftarrow \mathbf{H} \circ \frac{\mathbf{W}^T (\mathbf{V}\circ (\mathbf{X} + y))}{\mathbf{W}^T(\mathbf{V} \circ (\mathbf{WH} + y))} \\
    \mathbf{W} &\leftarrow \mathbf{W} \circ \frac{(\mathbf{V}\circ (\mathbf{X} + y))\mathbf{H}^T}{(\mathbf{V} \circ (\mathbf{WH} + y))\mathbf{H}^T}
\end{split}
\end{align}

When $\mathbf{X}$ is entirely non-negative no shift is required to ensure non-negativity 
and we can set $y = 0$. When these conditions are met 
these two update rules correctly recover the weighted vectorized update
rules (\ref{eq:update_weight}) derived in \cite{zhu_nonnegative_2016}. 
A proof that these update rules monotonically decrease  (\ref{eq:objective_shift}) is
provided in Appendix \ref{proof_shift}.

\begin{figure}[!t]
\centering
\includegraphics[width=\columnwidth]{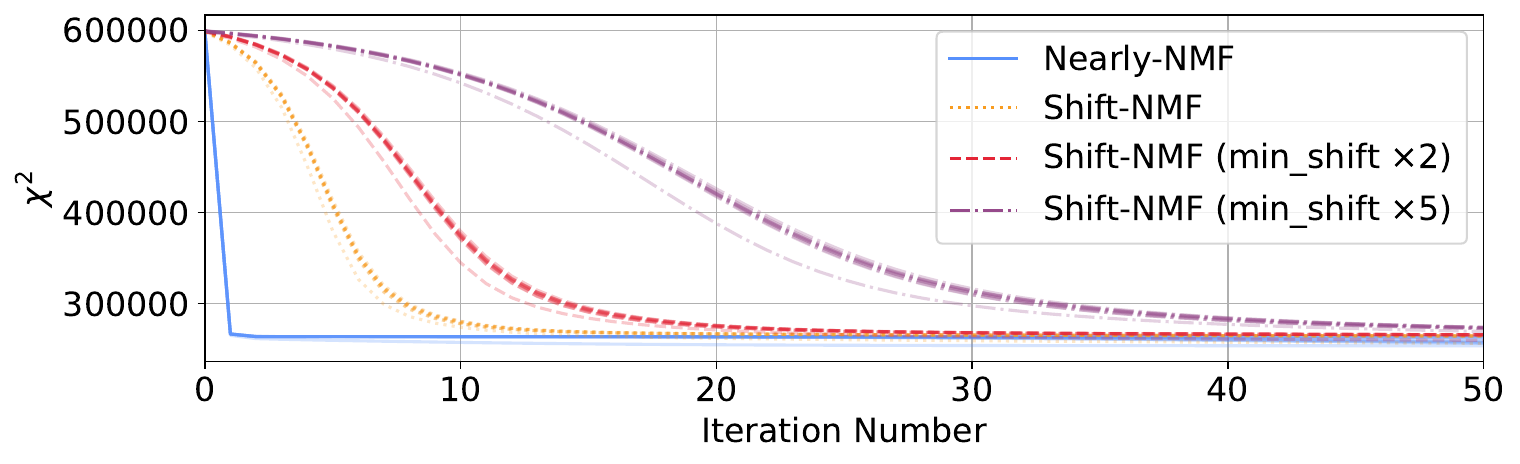}
\caption{Euclidean distance during training of two templates on the toy problem set out in the Introduction for the first 50 iterations, for a variety of different NMF algorithms. Nearly-NMF is plotted in solid blue, and Shift-NMF is plotted in dotted orange. In dashed red Shift-NMF was trained with a value of $y$ that is twice the minimum shift required, and in dash-dot purple $y$ was set to five times the minimum shift. Each test was 
rerun 10 times with different starting points, to remove the possibility of starting point bias. It is evident that increasing the shift value beyond the minimum slows the convergence of the $\chi^2$ value, while still training to comparable minimums given enough iterations.}\label{fig:toy_convergence}
\end{figure}

\subsection{Intermediate Steps: Non-scalar Shifts}
The update rules presented in (\ref{eq:update_shift}) have the caveat that the speed at which the objective $\chi^2_{\text{SHIFT}}$
is minimized is proportional to the magnitude of the shift $y$. In the limit where $y$ is much larger than any single element of $\mathbf{WH}$, the multiplicative
update factors for $\mathbf{W}$ and $\mathbf{H}$ (in this case, $\frac{\mathbf{W}^T (\mathbf{V}\circ (\mathbf{X} + y))}{\mathbf{W}^T(\mathbf{V} \circ (\mathbf{WH} + y))}$ for $\mathbf{H}$ and
$\frac{(\mathbf{V}\circ (\mathbf{X} + y))\mathbf{H}^T}{(\mathbf{V} \circ (\mathbf{WH} + y))\mathbf{H}^T}$ for $\mathbf{W}$) converge to 1.
This means that larger shift values decrease the convergence rate due to smaller updates per iteration.
An example of this is presented in Fig.~\ref{fig:toy_convergence}, where Shift-NMF was run on the same
toy dataset as before with increasingly large shift values of twice and five times the minimum scalar shift required but
with the same initial starting templates and coefficients. 
Also included on the plot is the convergence of Nearly-NMF, the derivation of which follows. Evidently, even though all three
Shift-NMF runs are converging to similar or the same local minimum, 
they are doing so at different rates dependent on the magnitude of the shift parameter $y$.

Since the rate of convergence is dependent on shift it is therefore optimal to consider using the 
smallest shift necessary in order to maximize the speed of convergence. 
The minimum possible shift would shift each pixel individually, and only by as much as necessary,
rather than shifting the entire dataset by a single scalar. To find this minimum possible
shift we will promote the necessary shift from a scalar shift
$y$ to a shift matrix $\mathbf{Y}$. For an arbitrary shift matrix $\mathbf{Y}$ the objective function is
\begin{equation} \label{eq:objective_nearly}
    \chi^2_\text{NEARLY} = ||\mathbf{V}^{1/2} \circ ((\mathbf{X} + \mathbf{Y}) - (\mathbf{WH} + \mathbf{Y})) ||^2
\end{equation}
which is once again identical to (\ref{eq:objective_weight}). We follow the same derivation as
in \ref{sec:shift-nmf} to produce the following update rules:
\begin{align}
\begin{split}\label{eq:h_x_prime}
    \mathbf{H} &\leftarrow \mathbf{H} \circ \frac{\mathbf{W}^T (\mathbf{V}\circ (\mathbf{X} + \mathbf{Y}_{\mathbf{H}}))}{\mathbf{W}^T(\mathbf{V} \circ (\mathbf{WH} + \mathbf{Y}_{\mathbf{H}}))}, \\
    \mathbf{W} &\leftarrow \mathbf{W} \circ \frac{(\mathbf{V}\circ (\mathbf{X} + \mathbf{Y}_{\mathbf{W}}))\mathbf{H}^T}{(\mathbf{V} \circ (\mathbf{WH} + \mathbf{Y}_{\mathbf{W}}))\mathbf{H}^T}.
\end{split}
\end{align}

For clarity we have added a subscript to the shift matrix $\mathbf{Y}$ for each update rule, as in
the most general case the shift matrix is not required to be identical for $\mathbf{H}$ and
$\mathbf{W}$ updates.


\subsection{Algorithm 2: Nearly-NMF} \label{sec:nearly-nmf}
Once we have the update rules (\ref{eq:h_x_prime}) we can derive the absolute minimum
possible shift, $\mathbf{Y}_{\text{min}}$, for quickest convergence. We will focus on $\mathbf{H}$, 
and the corresponding $\mathbf{Y}_{\mathbf{H},\text{min}}$, for the following derivation but the same
method applies to $\mathbf{W}$. In order to maintain the non-negativity of $\mathbf{H}$ at any
point we need only to ensure that the multiplicative update factor for $\mathbf{H}$ is (elementwise) non-negative:
\begin{equation} \label{eq:ineq_frac_nearly}
    \frac{\mathbf{W}^T (\mathbf{V}\circ (\mathbf{X} + \mathbf{Y}_{\mathbf{H}}))}{\mathbf{W}^T(\mathbf{V} \circ (\mathbf{WH} + \mathbf{Y}_{\mathbf{H}}))} \geq 0
\end{equation}

Since we expect the denominator of this inequality to all be entirely non-negative (verified below), we can drop the denominator from this inequality
entirely and only require
\begin{equation} \label{eq:ineq_nearly}
    \mathbf{W}^T (\mathbf{V}\circ (\mathbf{X} + \mathbf{Y}_{\mathbf{H}})) \geq 0
\end{equation}
to define our minimum shift $\mathbf{Y}_{\mathbf{H},\text{min}}$. Since $\mathbf{W}$ and $\mathbf{H}$ will change
at each iteration of the optimization, so too will the minimum shift matrix $\mathbf{Y}_{\mathbf{H},\text{min}}$. 
The minimum shift required to satisfy (\ref{eq:ineq_nearly}) can be found using
\begin{equation*}
\mathbf{W}^T (\mathbf{V} \circ \mathbf{Y}_{\mathbf{H},\text{min}}) = \begin{cases}
-\mathbf{W}^T (\mathbf{V} \circ \mathbf{X}),&{\text{if}}\ \mathbf{W}^T (\mathbf{V} \circ \mathbf{X}) < 0 \\ 
{0,}&{\text{otherwise,}} 
\end{cases}
\end{equation*}
which can be written 
\begin{equation} \label{eq:y_nearly}
\mathbf{W}^T (\mathbf{V} \circ \mathbf{Y}_{\mathbf{H},\text{min}}) = [\mathbf{W}^T (\mathbf{V} \circ \mathbf{X})]^-.
\end{equation}

 It is not necessary to directly solve
for the value of $\mathbf{Y}_{\mathbf{H},\text{min}}$ since it only appears in (\ref{eq:h_x_prime})
through the factor $\mathbf{W}^T (\mathbf{V} \circ \mathbf{Y}_{\mathbf{H}})$.
Furthermore, $\mathbf{W}^T (\mathbf{V} \circ \mathbf{X}) + [\mathbf{W}^T (\mathbf{V} \circ \mathbf{X})]^- =  
[\mathbf{W}^T (\mathbf{V} \circ \mathbf{X})]^+$. We now for completeness validate our assumption that
the denominator of (\ref{eq:ineq_frac_nearly}) is non-negative given this definition. Using (\ref{eq:y_nearly}) the denominator of (\ref{eq:ineq_frac_nearly}) is
\begin{equation*}
\begin{split}
    \mathbf{W}^T(\mathbf{V} \circ (\mathbf{WH} &+ \mathbf{Y}_{\mathbf{H}})) \\
    &= \mathbf{W}^T(\mathbf{V} \circ (\mathbf{WH})) + [\mathbf{W}^T (\mathbf{V} \circ \mathbf{X})]^-
\end{split}
\end{equation*}
which is correctly non-negative given non-negative initialization of $\mathbf{W}$ and $\mathbf{H}$.

Using these derivations we can rewrite the $\mathbf{H}$ update in (\ref{eq:h_x_prime}) without
any reference to the shift matrix $\mathbf{Y}_{\mathbf{H}}$. A
similar derivation can show that 
\begin{equation}
    (\mathbf{V} \circ \mathbf{Y}_{\mathbf{W}, \text{min}}) \mathbf{H}^T  = [(\mathbf{V} \circ \mathbf{X}) \mathbf{H}^T]^-,
\end{equation}so our final update rules can be written:
\begin{align}
\begin{split}\label{eq:update_nearly}
    \mathbf{H} &\leftarrow \mathbf{H} \circ \frac{[\mathbf{W}^T(\mathbf{V}\circ\mathbf{X})]^+}{\mathbf{W}^T(\mathbf{V}\circ(\mathbf{WH})) + [\mathbf{W}^T(\mathbf{V}\circ\mathbf{X})]^-} \\
    \mathbf{W} &\leftarrow \mathbf{W} \circ \frac{[(\mathbf{V}\circ\mathbf{X})\mathbf{H}^T]^+}{(\mathbf{V}\circ(\mathbf{WH}))\mathbf{H}^T + [(\mathbf{V}\circ\mathbf{X})\mathbf{H}^T]^-}
\end{split}
\end{align}

An important last part of the derivation is to underscore that in general 
$[\mathbf{W}^T(\mathbf{V}\circ\mathbf{X})]^+ \neq \mathbf{W}^T[(\mathbf{V}\circ\mathbf{X})]^+$, and
correspondingly for the other splits.

We call this method Nearly-NMF, as it is ``nearly'' non-negative,
in that it keeps the non-negative constraints on $\mathbf{W}$ and $\mathbf{H}$, while removing
the non-negative constraint on the data $\mathbf{X}$. 
In the limit where $\mathbf{X}$ contains no negative elements, the two terms
 $[\mathbf{W}^T(\mathbf{V}\circ\mathbf{X})]^+ \rightarrow \mathbf{W}^T(\mathbf{V}\circ\mathbf{X})$ and $[(\mathbf{V}\circ\mathbf{X})\mathbf{H}^T]^+ \rightarrow (\mathbf{V}\circ\mathbf{X})\mathbf{H}^T$, with $[\mathbf{W}^T(\mathbf{V}\circ\mathbf{X})]^- \rightarrow 0$ and $[(\mathbf{V}\circ\mathbf{X})\mathbf{H}^T]^- \rightarrow 0$ following, and we
once again correctly recover the weighted update rules (\ref{eq:update_weight}).
A proof that these update rules monotonically decrease (\ref{eq:objective_nearly})
with the given constraints is provided in Appendix \ref{sec:proof_nearly}.

\subsection{Theoretical Comparison Between Shift-NMF and Nearly-NMF}
Shift-NMF is elegant in its simplicity, both conceptually and implementation-wise. However,
we want to underscore that Nearly-NMF is superior in all other ways, and its minimum 
shift optimizations allow it to converge to a similar minima of the objective function
in fewer iterations than Shift-NMF. Emperically we observe that
Nearly-NMF update rules at worst minimizes the objective (\ref{eq:objective_weight}) 
the same amount at each iteration as Shift-NMF for the same input templates and coefficients,
and in general minimizes much faster than Shift-NMF, but we do not provide a formal proof
of this statement. 

\section{Numerical Examples}
\subsection{Toy Problem}
\begin{figure}[!t]
\centering
\includegraphics[width=\columnwidth]{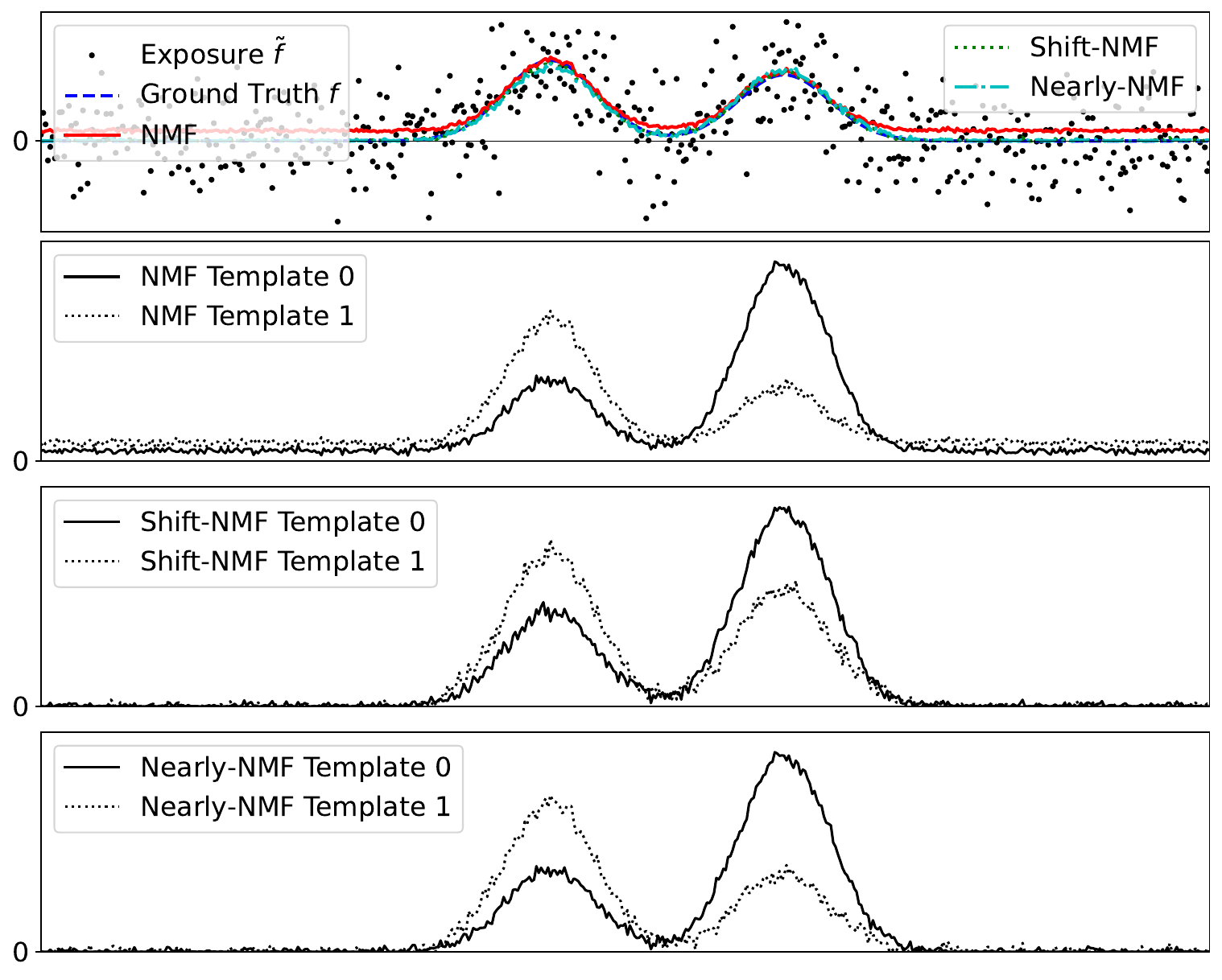}\label{toy_with_negs}
\caption{Results of two NMF templates generated on a toy example, for each of the three algorithms: regular weighted NMF, Shift-NMF, and Nearly-NMF. See Introduction for details of toy example. Top panel shows one representative exposure from the set of 500 in dots, with the noiseless truth and the template based reconstructions overplotted in varying styles. Note that the reconstructions from two methods presented in this paper, Shift-NMF and Nearly-NMF, are nearly indistinguishable from the truth, and lie directly on top of it. The next three panels shows the two raw templates, scaled so the maximum value is 1 but preserving the relative scale between the two templates. Shift-NMF and Nearly-NMF templates correctly go to zero on the edges, whereas weighted NMF has a vertical offset.}
\label{fig:toy_with_negs}
\end{figure}

As a first test that Shift-NMF and Nearly-NMF correctly handle the negative noise, we will return to the
toy problem set up in the Introduction. Using the same dataset, the results of running Shift-NMF and Nearly-NMF
are plotted in Fig.~\ref{fig:toy_with_negs}. Both Shift-NMF and Nearly-NMF correctly generate templates
that account for the negative components of the noise, and do not present the positive offset
that weighted or standard NMF does. 

\subsection{Simulated Data}
\subsubsection{Data Generation}
For a more realistic example, we turn to an astronomy-motivated case ---
building NMF templates to model quasar spectra (flux vs. wavelength of light).
Quasars are among the brightest objects in the universe, visible to billions of
light-years away using large modern research telescopes.
As the light travels from the quasar to earth, the expansion of the universe
causes a multiplicative shift in the wavelengths of the light by a factor of
$(1+z)$ where $z$ is the ``redshift'' of the quasar.  More distant quasars have
larger redshifts due to the light experiencing more expansion of the
universe between the time it leaves the quasar and when it arrives at earth.
Templates of quasar spectra are used to model newly observed
spectra to confirm their identification and measure their redshift.
\cite{bolton_sdss_spec1d_2012} used 568 quasar spectra to build Principal Component Analysis (PCA)
templates for the Baryon Oscillation Spectroscopic Survey (BOSS);
\cite{brodzeller_qsotemplates_2023} used a much larger set of 207\,956 quasar spectra
from the extended Baryon Oscillation Spectroscopic Survey (eBOSS) \cite{lyke_dr16q_2020}
to build new PCA templates for the Dark Energy Spectroscopic Survey
\cite{desi_sv_overview_2023, guy_desi_pipeline_2023}.
Although PCA algorithms such as
EMPCA \cite{bailey_empca_2012} (used by \cite{brodzeller_qsotemplates_2023}) and \cite{delchambre_wpca_2015} accommodate noisy, negative,
and missing data, a downside of PCA templates is that they can be overly
flexible in overfitting noise, since they don't have the constraint that the reconstructed 
fit to the data is strictly positive.  NMF templates could provide improved
results by eliminating a class of unphysical false positives in the fits of templates to data.

To have a ground truth for comparison in this study, we used simqso \cite{mcgreer_simqso_2021}
to generate 200\,000 simulated quasar spectra with random uniform redshifts between 0 and 4.
We use 130\,000 of these spectra as a training sample, and hold out the remaining 70\,000 as a validation
sample.
The generated spectra have logarithmically spaced wavelengths $\lambda$
from 360--1000~nm with a spacing of $\Delta \log{\lambda} = 10^{-4}$,
approximately matching the wavelength coverage of the BOSS quasar sample \cite{dawson_sdss-iv_2016}.
In order to simulate Poisson statistics we convert the simulated flux to a photon
estimate 
and use the scaled noiseless value in each wavelength bin
as the mean for a random Poisson draw.
We use that Poisson value as the spectral value for that wavelength. We add zero-mean Gaussian noise to these
resultant spectra, 
such that the expected average SNR of the Gaussian noise is around 1.5
\footnote{Real astronomical spectra have an even more complex noise model due to
wavelength-dependent instrument throughput and backgrounds, but the level of complexity used
in the simulated data for this study is sufficient to demonstrate the performance of the
Shift- and Nearly-NMF methods.}. This procedure
is identical to that of (\ref{eq:simple_test}), with the difference that we no longer
randomly choose $\sigma_{\text{noise}}^2$, instead determining it from the data
to reach our target SNR.

Observed spectra cover a fixed wavelength range in the ``observer-frame'',
however, due to the redshift effect these correspond to a different wavelength
range per quasar in their source ``rest-frame''.  Thus building an NMF model that
covers all wavelengths is not just a noisy-data problem, but also a missing-data
problem since no individual spectrum covers all wavelengths of interest.
To build templates, we
de-redshift all simulated quasars onto common wavelength grid in their rest-frame,
using weights of 0 for any missing data due to any wavelengths that cannot
be measured in the observer-frame.
This produces an input dimensionality, $d$, of 11\,400 pixels spanning all wavelengths.
Finally, we re-normalize the median flux of each quasar to match the median
flux over the same rest-frame wavelength range of a common reference quasar
so that variations are due to spectral diversity rather than overall amplitude fluctuations. 
We also correspondingly rescale the variance using the same scaling factor $s$ used to rescale
the data:
\begin{equation}
    \tilde{\sigma}^2 \leftarrow s^2 * \tilde{\sigma}^2
\end{equation}

\begin{figure}[!t]
\centering
\includegraphics[width=\columnwidth]{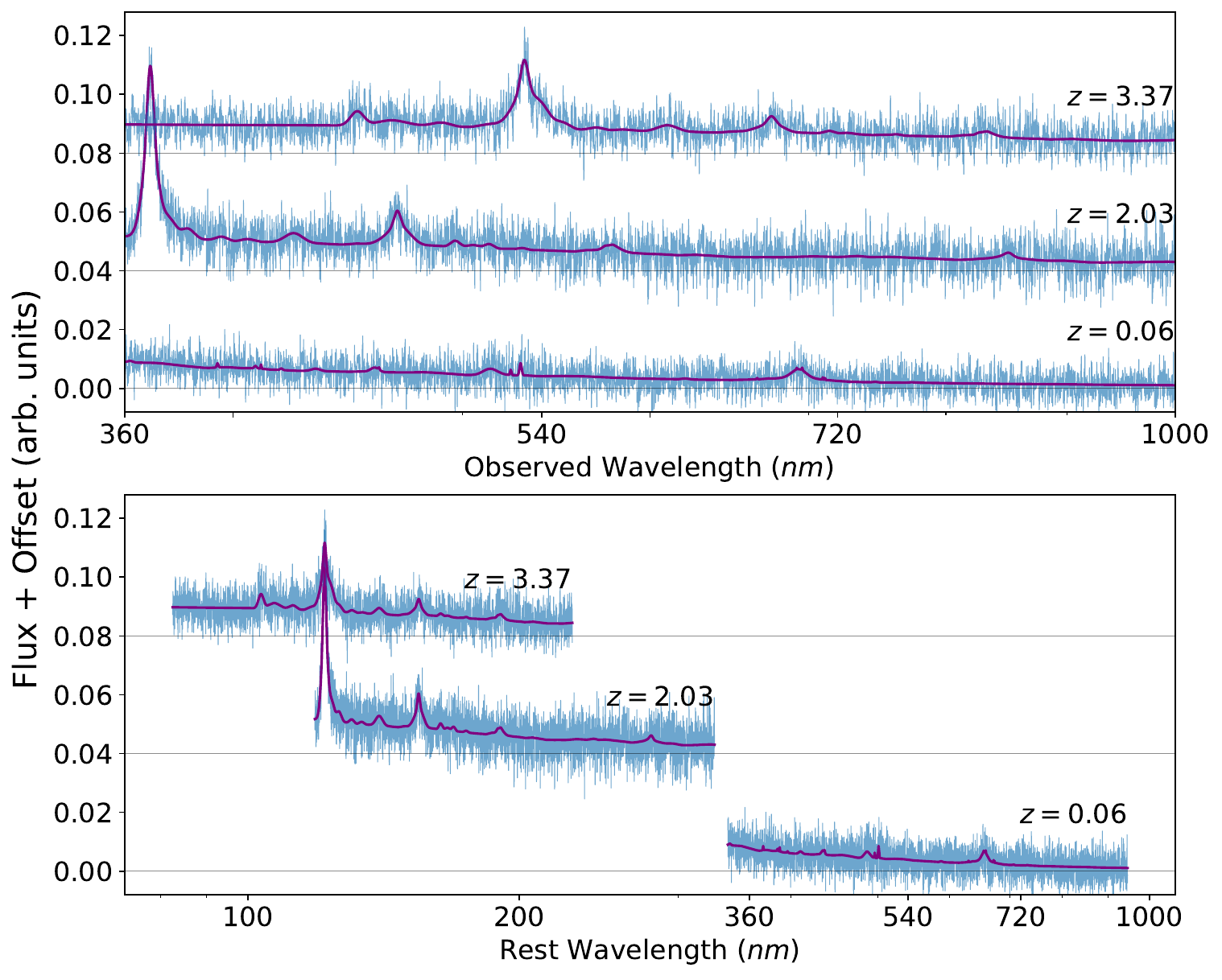}
\caption{An example of the process used to generate the quasar dataset. In both panels three unnormalized spectra are offset vertically from each other
by a constant value 0.04. In the upper panel, three different noisy spectra are plotted in light blue
in the observed frame, as the simulated instrument would record them, with their noise-free base spectra
overplotted in purple.
The three spectra are annotated by their redshift value at the far right of the plot.
Note in the noisy data the high prevalence of negative values due to the noise. In the lower panel we demonstrate the spectra as they are used in both
Shift-NMF and Nearly-NMF, now in the rest frame. It is evident that the 
spectra cover different amounts of the wavelength grid, with none of them covering
the entirety of the fitting space. The spectra plotted here are not renormalized, which is done before fitting.}
\label{fig:example_qsos}
\end{figure}

An example of this process is shown in Fig. \ref{fig:example_qsos}. In the upper panel
are three simulated quasars in the observed frame, with the noise-free truth plotted over the 
noisy simulated exposure. In the bottom panel these quasars are shown in the rest-frame, plotted
on the common wavelength grid. 
It is evident that none of the individual quasar spectra cover the entire
wavelength range to be modeled, and every wavelength is only covered by a subset
of the input training spectra, with no individual pixel covered by every spectra.  
Before fitting the templates, we trim the common wavelength grid to avoid edge effects due to
low numbers of quasars in edge pixels. We trim 50 pixels off the low end and 300 pixels off the high
end to fit a total range of $11050$ pixels. 
This presents a realistically challenging dataset
for NMF modeling, including $61.1\%$ of data being ``misssing,''
$11.4\%$ of non-missing data being negative (due to noisy ``observations''
of an underlying strictly positive signal), 
variable noise, and a $11\,050 \times 130\,000$ dataset to solve
for a set of NMF templates modeling all wavelengths.
Detailed studies on NMF's performance in the presence of missing data 
and its performance on corresponding data imputation tasks specifically 
in astrophysics have been carried out in \cite{zhu_nonnegative_2016, ren_using_2020}.
In this work we will focus only on Nearly-NMF's capability to generate templates given the 
noted data challenges.

\subsubsection{Templates}
\begin{figure}[!t]
    \centering
    \includegraphics[width=\columnwidth]{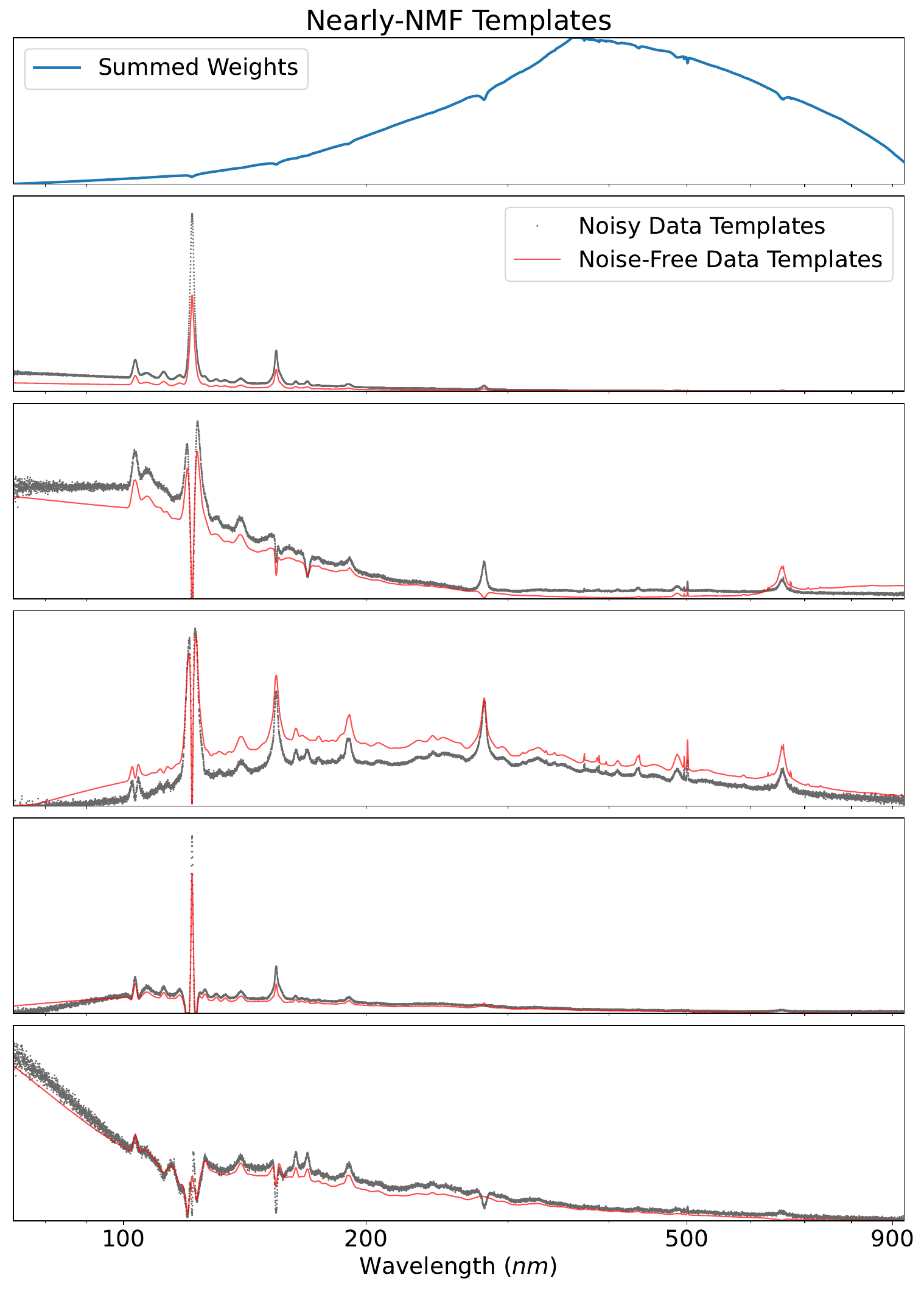}
    \caption{The top panel shows the sum of the inverse variance weights in each pixel over the wavelength region covered by the templates. The next 5 panels show each of the 5 Nearly-NMF templates, generated on both the noisy and noise-free datasets. The noise-free templates are plotted as a solid line in red, while the noisy templates are dotted in light gray. Templates are plotted on their logarithmic grid, with a logarithmic scaling on the x-axis. The noisy templates have still recovered most of the same features
    present in the noise-free templates, and have good agreement even though the
    noisy data has a significant amount of negative values. The regions where the templates are noiseiest correspond with the same regions where the sum of the weights of the training data is low. Note that NMF based algorithms do not produce unique factorizations, and these are only one of many possible factorizations.}
    \label{fig:simqso_nearly}
\end{figure}

For this simulated test we focused on Nearly-NMF due to its faster convergence than Shift-NMF, and
generated 5 templates on both the noisy datasets as well as the noise-free, true, underlying spectra. 
We will use the templates generated on the true spectra to validate how well the algorithm handles 
both the noise and the introduced negativity.
In order to generate templates ordered such that subsequent templates approximately 
explain less variance than the
previous template, similar to that of PCA, we generate the
templates sequentially. The sequential NMF method is detailed further in \cite{zhu_nonnegative_2016}
and \cite{ren_non-negative_2018}, but we will provide
a summary. We start by generating one template from a linear starting
point with random uniform coefficients for each exposure. 
After that template is trained for a smaller number of iterations ($\sim 50$) we stop and train
two templates, with the first template initialized to the previously trained template and the
new template initialized to a linear start, with a corresponding trained coefficient and random 
coefficient respectively for this set of templates. We do not hold the previously trained template
fixed, allowing it to continue to be trained.
We repeat this sequential method until we have 5 templates.
The 5 templates are then refined through an additional 1000 iterations of updates. 
The final number of iterations, 1000, is chosen to ensure convergence to a local minima template.
The final Nearly-NMF templates generated in this test are shown in 
Fig. \ref{fig:simqso_nearly}.
All templates are plotted in the order they were generated in the sequential methodology explained
prior. 
Note that although the sequential method helps reduce the sensitivity of the final templates to 
their initialization points, it does not completely remove this effect, and the templates we
show are only one of many possible answers.

For all cases it can be seen that Nearly-NMF has accounted for the noise
in the input spectra, and has recovered templates that incorporate many of the same details
that the noise-free templates capture. Both sets of templates are also correctly non-negative, and
feature no negative values. Nearly-NMF has
correctly accounted for the zero-mean property of the noise and 
has produced templates that appear similarly close
to zero as the noise-free templates. 

The increased noise in the noisy templates in the low wavelength and high wavelength regions,
seen visually in  Fig. \ref{fig:simqso_nearly} as a wider grey band in the noisy templates
at the edges of the figure as compared to the center wavelengths,
is to be expected. Due to the non-linear nature of de-redshifting spectra,
the number of spectra and the corresponding
summed weight in each pixel is not uniform.
A plot of the summed weight in each pixel is shown in the 
top panel of Fig. \ref{fig:simqso_nearly}, where it is evident
that the noisier regions of the templates correspond with lower total summed weight in that pixel.

\subsubsection{Validation of Templates}
\begin{figure}[!t]
    \centering
    \includegraphics[width=\columnwidth]{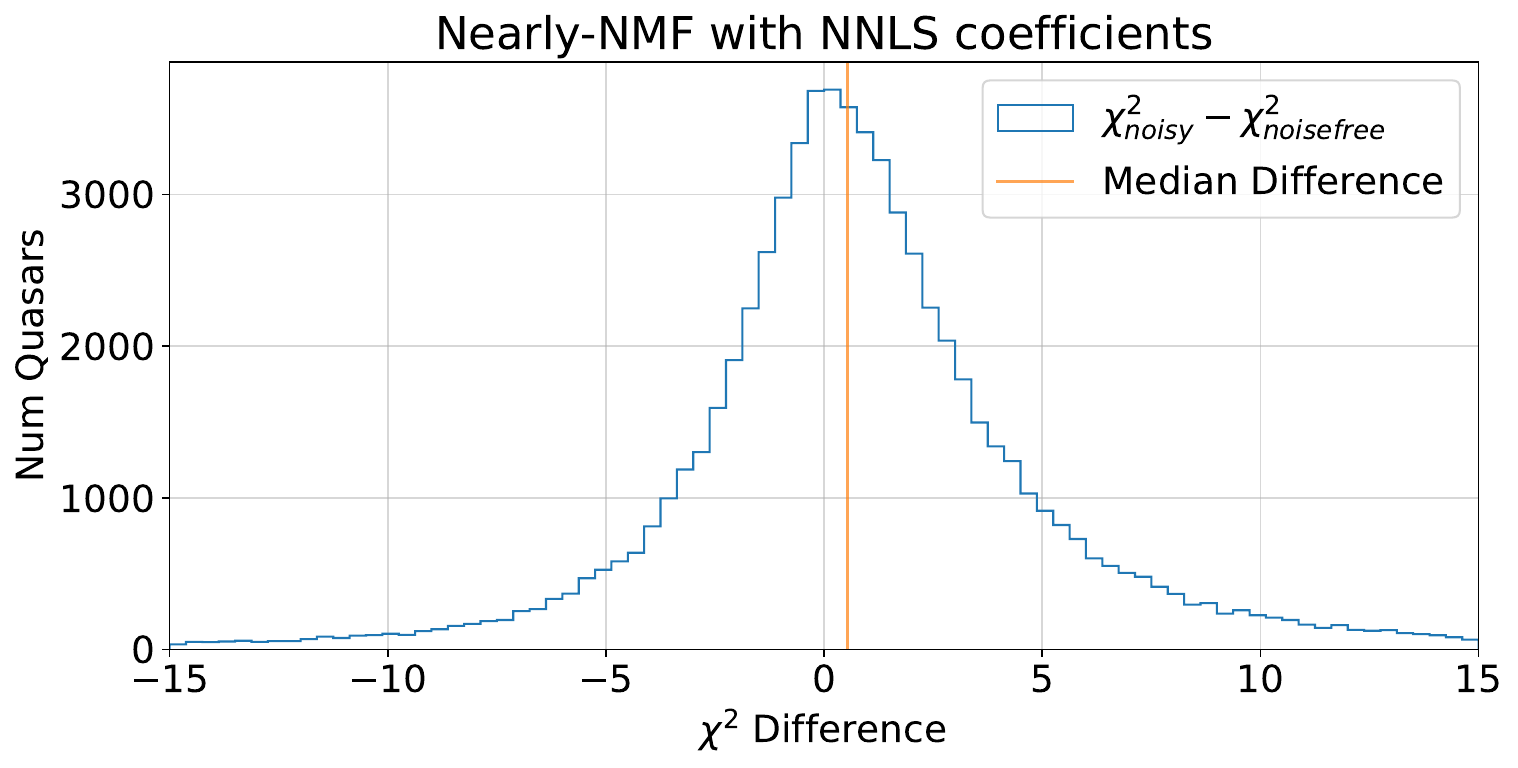}
    \caption{Comparison between the $\chi^2$ values from fitting the $70\,000$ validation dataset by both the noisy and noise-free templates. A vertical line at about $0.544$ indicates the median difference,
    compared to the typical absolute $\chi^2 \sim 4440$. 
    A small positive offset means that the noisy templates fit the data slightly worse, 
    but its proximity to zero indicates that they are comparably good fits.}
    \label{fig:chi_2}
\end{figure}

In order to validate that the templates generated on the noisy dataset 
are comparably good to the templates
generated on the noise-free truth dataset, we fit both sets of templates to the holdout $70\,000$
validation dataset of quasars. 
For this test we use a non-negative least squares solver
provided by scipy (\texttt{scipy.optimize.nnls}) \cite{2020SciPy-NMeth} 
to find coefficients for each template 
to remove the impact of picking a bad starting point for the fit.
Once we fit both sets of templates to the validation dataset, we calculate their
$\chi^2$ values using (\ref{eq:objective_nearly}), and then make a histogram of 
the difference between the $\chi^2$ of the noise-free and noisy template fits. 
These results are shown in Fig. \ref{fig:chi_2}. A vertical line denotes the median 
difference between the noise-free and noisy fits, at about $0.544$, in comparison to the
typical absolute $\chi^2 \sim 4440$.
The slight positive shift of the data indicates that the noisy templates
fit the data slightly worse than the noise-free templates, but the remarkably sharp peak near
0 indicates that even though one set of templates are generated on noisy data with a non-trivial
amount of negative values, they are fitting the dataset comparably well as templates generated
on the underlying noise-free truth, which has no noise and is entirely non-negative.

\begin{figure}[!t]
\centering
\includegraphics[width=\columnwidth]{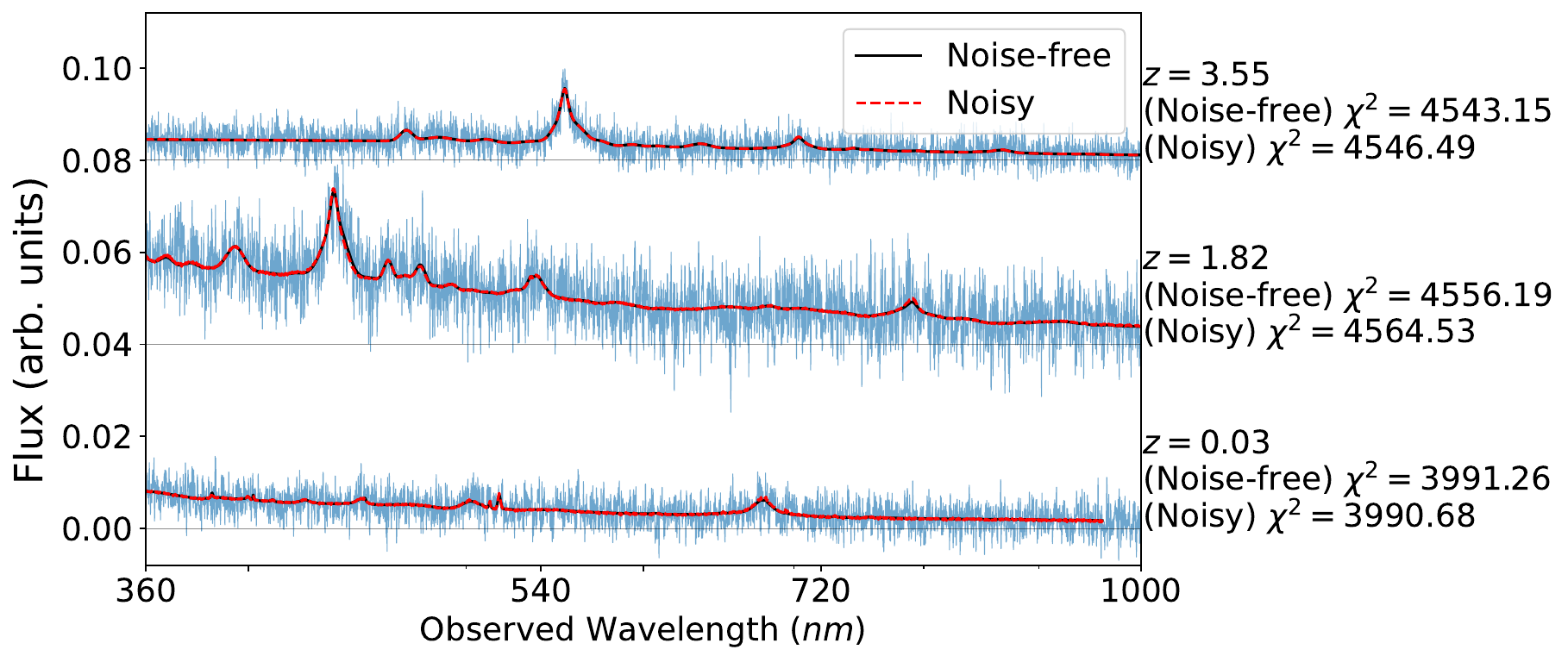}
\caption{Three quasars from the noisy validation set plotted in the observed frame with their
Nearly-NMF reconstructions overplotted, with the reconstruction from the noise-free templates
in dot dash green and the reconstruction from the noisy templates in dashed red. Both
reconstructions are comparable and nearly identical, validated by their similar Euclidean distances from the data,
annotated at right. It is notable that the two reconstructions, despite one being derived
from templates trained on noisy data, are remarkably similar. Note also some missing pixels in the 
reconstruction of the bottom spectra, due to the clipping of the fitting region to avoid edge effects.
In this plot we are plotting the non-renormalized spectra,
as in Fig. \ref{fig:example_qsos}, with the reconstructions rescaled to match.}
\label{fig:qsos_with_recon}
\end{figure}

An example of how closely matched these reconstructions between that of the noisy and noise-free templates
is given in Fig. \ref{fig:qsos_with_recon}. Plotted in the figure are three quasars of similar redshift
to those plotted in Fig. \ref{fig:example_qsos}, however these quasars are drawn from the validation set
and were not included in training. 
Their fits from the noise-free and noisy templates are both overplotted, and are similar 
but not identical to each other. 

\section{Computational Properties of the Algorithms}
\begin{figure*}[t]
    \centering
    \includegraphics[width=\textwidth]{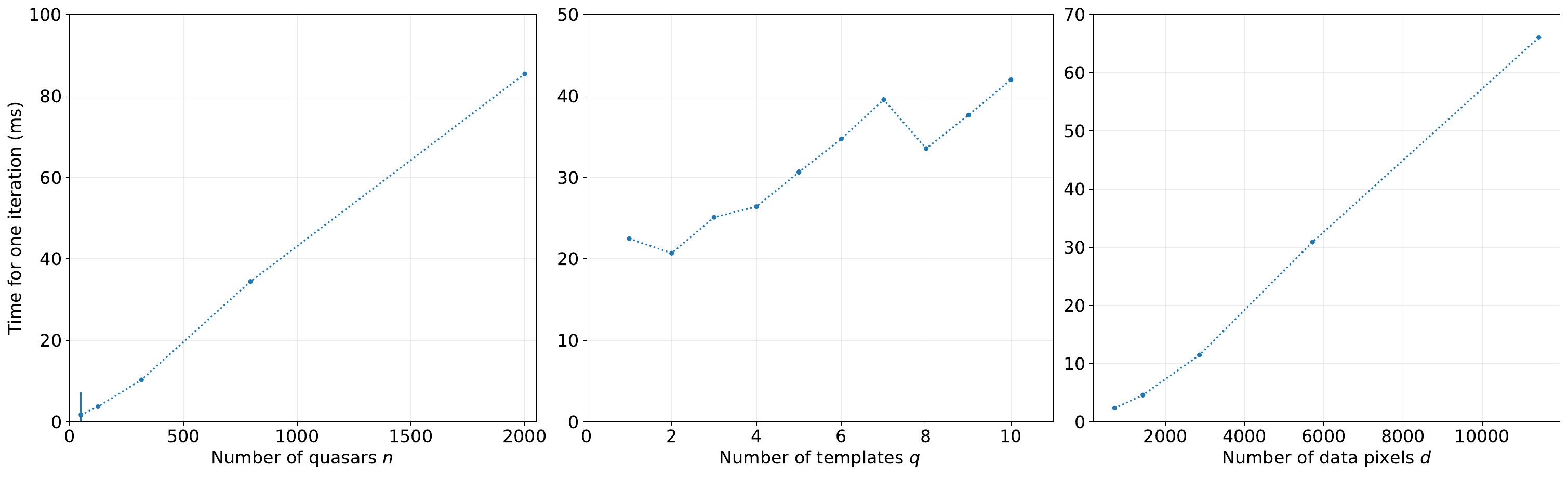}
    \caption{Results of three different timing runs for Nearly-NMF. In each plot we hold the other two variables fixed. The fixed values for the variables not varying in each plot are $N=750$, $q=5$ and $d= 5700$. Each point is the median time to complete a single iteration of updates, taken from 100 subsequent iterations. Error bars are plotted as the standard deviation of these 100 iterations. Note that in almost all cases the error bars are smaller than the point size, representing considerable stability in timing.}
    \label{fig:timing}
\end{figure*}

In this section we will investigate some basic computational properties of the Nearly-NMF algorithm.
In theory we would expect both algorithms to scale, in big-$O$ notation, roughly
in terms of $O(n\times d\times q\times n_{iter})$, where $n, d, q$ are the variables set out in the Introduction, 
and $n_{iter}$ is the chosen number of iterations to run the update rules. To test this
we set up individual parameter sweeps for the three values of interest ($n, d, q$) 
and time the amount of time it takes to do one single iteration of both $\mathbf{H}$ and $\mathbf{W}$ Nearly-NMF
update rules. Each test is run for 100 iterations, and we plot the median values of these
100 runs with error bars equal to the standard deviation of the runs. For these
timing runs we force the algorithm to use only a single thread but some speedup, especially
at higher values of $d$ and $n$, can be achieved by allowing it to use more threads. Additionally,
although we implemented and used a GPU version of both algorithms for generating the templates
in Fig. \ref{fig:simqso_nearly}, we used the CPU implementation
for the following timing tests. These tests
were run on a dual AMD EPYC 7763 CPU system with 512 GB of RAM, corresponding to a single node
on the National Energy Research Scientific Computing Center's (NERSC) Perlmutter system.
The results of the timing runs are included in Fig. \ref{fig:timing}.
Specific details for each metric's timing runs are as follows:

\begin{itemize}
    \item For $n$, we train 5 templates, and use half of the original dimensionality of the dataset presented
in this paper ($d=5700$). We take a logarithmically spaced number quasars from 50 up to 2000 and at
each point use a randomly selected subsample of the quasar superset generated for this paper.

\item For $q$ we use a fixed value of 750 quasars, and half of the original dimensionality of the dataset presented
in this paper. We vary the number of templates trained from 1 up to 10.

\item For $d$ we use a fixed value of 750 quasars and train 5 templates. We rebin the input dataset
from the original to varying sizes such that each subsequent rebinning is half
the dimensionality of the previous.
\end{itemize}

We have also confirmed that the scaling relations hold true for Shift-NMF, with 
per-iteration timings similar
to those reported for Nearly-NMF in Fig. \ref{fig:timing}, with the same scaling relations, but
do not explicitly plot them.
As expected for all three variables, the scaling relations are approximately linear.
For the case of number of templates $q$, we note that the scaling does not intercept at time=0
since the update rules include terms with element-wise multiplication by $\mathbf{V}$,
$(\mathbf{V} \circ \mathbf{X})$ and $(\mathbf{V} \circ \mathbf{W} \mathbf{H})$,
which have the same dimensionality regardless of $q$.

\section{Conclusion}
In this paper we have presented two new versions of non-negative matrix factorization, Shift-NMF
and Nearly-NMF,
that can adequately handle noisy data with negative values introduced by noise while still obeying the non-negativity constraint on the coefficients and
templates. Both of these algorithms are robust to the noise as well as missing data 
and have straightforward update rules akin to those of \cite{lee_algorithms_2000}. We have also shown
that in a realistic test with large amounts of missing and negative data that Nearly-NMF can recover
templates giving equally as good fits as templates generated on the noise-free truth.

Both of these methods were developed in an astronomical context, but their applications are 
not limited to only this regime. Non-negative matrix factorization has been applied
to a variety of other fields where negative data can arise either naturally
in the data or as part of a data processing step. 
Some examples include \cite{mccarty_blind_2020} for NMR spectroscopy,
\cite{boashash_chapter_2016} for brain pattern analysis in neuroscience or \cite{kim_subsystem_2003} for bioinformatics.
All of these fields might benefit from NMF style algorithms that can handle some negative
values in the input data, when the negative data is expected to result from zero-mean noise.

We release our implementations of the Shift-NMF and Nearly-NMF algorithms for public use at
\url{https://github.com/dylanagreen/nearly_nmf}. 
This code is written in Python, is optimized for both CPU use through numpy \cite{harris_array_2020}, and GPU use through cupy \cite{cupy_learningsys2017}, and was used
to train the templates plotted in this paper. Included
within the linked repository is all code necessary to reproduce the plots and analysis in this paper. 

\appendix{}\label{sec:proofs}

This appendix includes proofs that the update rules provided in the main 
text of the paper monotonically decrease their respective $\chi^2$ objectives as well as a proof
that Nearly-NMF minimizes its objective at least as fast as Shift-NMF. 
In this section we introduce the convention that $[\mathbf{\boldsymbol{\alpha}}]^{diag}$ 
refers to a diagonal matrix constructed by placing the elements of
the vector $\boldsymbol{\alpha}$ along the diagonal of a matrix with all other elements set to zero.

\subsection{Shift-NMF}\label{proof_shift}
To prove the Shift-NMF update rules monotonically decrease their $\chi^2$ objective we will start
with $\mathbf{H}$.
We can rewrite the $\mathbf{H}$ update in (\ref{eq:update_shift}) in column-wise form as 
\begin{equation} \label{eq:h_shift_column}
    \mathbf{H}_{,j} \leftarrow \mathbf{H}_{,j} \circ \frac{\mathbf{W}^T (\mathbf{V}_{,j}\circ (\mathbf{X}_{,j} + y))}{\mathbf{W}^T(\mathbf{V}_{,j} \circ (\mathbf{W}\mathbf{H}_{,j} + y))},
\end{equation}
and then use the following transforms that mirror those from \cite{zhu_nonnegative_2016}:
\begin{align}
\begin{split}
    \hat{\mathbf{X}}_{,j} &= \mathbf{V}_{,j} \circ (\mathbf{X}_{,j} + y), \\
     \hat{\mathbf{W}} &= [\mathbf{V}_{,j}]^{diag} \mathbf{W}, \\
      \hat{\Phi}_{,j} &= y\mathbf{V}_{,j},
\end{split}
\end{align}
to rewrite  (\ref{eq:h_shift_column}) as
\begin{equation}
    \mathbf{H}_{,j} \leftarrow \mathbf{H}_{,j} \circ \frac{\mathbf{W}^T (\hat{\mathbf{X}}_{,j})}{\mathbf{W}^T((\hat{\mathbf{W}}\mathbf{H}_{,j} + \hat{\Phi}_{,j})}.
\end{equation}

This equation matches the form of  (41) in \cite{tang_nonnegative_2012}, 
which is already proven to monotonically decrease the objective function. Therefore 
our H update rule in (\ref{eq:update_shift}) follows the same conclusion. 
A similar breakdown of the objective function relative to the rows of $\mathbf{W}$ can be worked through, 
and the same proof from \cite{tang_nonnegative_2012} can be applied to 
prove the same result for the W update rule.
 
\subsection{Nearly-NMF} \label{sec:proof_nearly}
In order to prove that the $\chi^2$ given in (\ref{eq:objective_nearly}) and identically
(\ref{eq:objective_weight}) is non-increasing under the update rules (\ref{eq:update_nearly}) we will follow the methodology of \cite{lee_algorithms_2000} and construct auxiliary functions 
$F$. We then prove that the update rules
are global minima of the auxiliary functions. The auxiliary functions 
that we use in our proof will fulfill two properties:
\begin{align} \label{eq:auxiliary_def}
\begin{split}
    F(\mathbf{A}, \Tilde{\mathbf{A}}) &\geq J(\mathbf{A}),\\
    F(\mathbf{A}, \mathbf{A}) &= J(\mathbf{A}),
\end{split}
\end{align}
where in our case $\mathbf{A}$ is either $\mathbf{W}$ or $\mathbf{H}$ and  $\Tilde{\mathbf{A}}$ is
any arbitrary matrix of the same shape as $\mathbf{A}$. The definition of $J(\mathbf{A})$
will depend on whether $\mathbf{A} = \mathbf{W}$ or $\mathbf{A} = \mathbf{H}$. We will 
outline this proof for $\mathbf{H}$ in the unweighted case, so we will use
\begin{equation}
    J(\mathbf{A}) = ||\mathbf{X} - \mathbf{WA} ||^2,
\end{equation}
and then show that under a simple transform, as in Section \ref{proof_shift},
the weighted case (\ref{eq:objective_weight}) fulfills the same proof. To start, it is useful to rewrite the 
unweighted objective (\ref{eq:objective}) in scalar form as
\begin{equation}\label{eq:objective_nearly_scalar}
\begin{split}
    J = \sum_{ij} \biggl( &-2(\sum_{k}[\mathbf{X}_{ij}\mathbf{W}_{ik}]^+\mathbf{H}_{kj}) \\&+ 2(\sum_{k}[\mathbf{X}_{ij}\mathbf{W}_{ik}]^-\mathbf{H}_{kj}) \\&+ \sum_{kl}\mathbf{H}_{kj}\mathbf{W}_{ik}\mathbf{W}_{il}\mathbf{H}_{lj} \biggr),
\end{split}
\end{equation}
where it is important to note that we have dropped the term $\sum_{ij}\mathbf{X}_{ji}\mathbf{X}_{ji}$
since minimizing the objective with respect to either $\mathbf{H}$ or $\mathbf{W}$ is independent of this term. Using the fact that for any scalar 
$a, b > 0$, $a \leq (a^2 + b^2) / 2b$, and for any non-negative vectors $\boldsymbol{\alpha, \beta}$ and non-negative matrix $\mathbf{P}$, $\sum_{ij} \boldsymbol{\alpha}_i\boldsymbol{\alpha}_j\mathbf{P}_{ij} \leq \sum_{ij} \boldsymbol{\alpha}_i^2 \frac{\boldsymbol{\beta}_j}{\boldsymbol{\beta}_i}\mathbf{P}_{ij}$ from \cite{blanton_k_2007}, 
we construct the auxiliary function
\begin{equation}
\begin{split}
    F(\mathbf{H}, \Tilde{\mathbf{H}}) = &-2\sum_{ijk}[\mathbf{X}_{ij}\mathbf{W}_{ik}]^+\mathbf{H}_{kj} \\&+ \sum_{ijk}[\mathbf{X}_{ij}\mathbf{W}_{ik}]^-\frac{\mathbf{H}_{kj}^2 + \Tilde{\mathbf{H}}_{kj}^2}{\Tilde{\mathbf{H}}_{kj}} \\ &+ \sum_{ijkl}\mathbf{W}_{ik}\mathbf{W}_{il}\mathbf{H}_{lj}^2\frac{\mathbf{\Tilde{H}}_{kj}}{\mathbf{\Tilde{H}}_{lj}}
\end{split}
\end{equation}

This function fulfills the criteria $F(\mathbf{H}, \mathbf{H}) = J(\mathbf{H})$, and given the above two
properties it also follows that $F(\mathbf{H}, \Tilde{\mathbf{H}}) \geq J(\mathbf{H})$
holds for any arbitrary $\mathbf{H} \geq 0$ and $\Tilde{\mathbf{H}} \geq 0$. Next we will prove that
$F(\mathbf{H}, \Tilde{\mathbf{H}})$ is convex. Proving the convexity
of  $F(\mathbf{H}, \Tilde{\mathbf{H}})$ necessitates the second derivative, allowing us to 
compute the first derivative along the way:
\begin{equation} \label{eq:df_dh}
\begin{split}
    \frac{\partial F(\mathbf{H}, \Tilde{\mathbf{H}})}{\partial \mathbf{H}_{ab}} = &-2\sum_{j}[\mathbf{X}_{ib}\mathbf{W}_{ia}]^+ \\&+ 2\sum_{i}[\mathbf{X}_{ib}\mathbf{W}_{ia}]^-\frac{\mathbf{H}_{ab}}{\Tilde{\mathbf{H}}_{ab}} \\ &+ 2\sum_{ik}\mathbf{W}_{ik}\mathbf{W}_{ia}\mathbf{H}_{ab}\frac{\mathbf{\Tilde{H}}_{kb}}{\mathbf{\Tilde{H}}_{ab}}
\end{split}
\end{equation}
\begin{equation} \label{eq:d2f_dh2}
\begin{split}
    \frac{\partial^2 F(\mathbf{H}, \Tilde{\mathbf{H}})}{\partial \mathbf{H}_{ab}\partial \mathbf{H}_{cd}} = 2 \delta_{ac}\delta_{bd} \biggl(&\sum_{i}\frac{[\mathbf{X}_{ib}\mathbf{W}_{ia}]^-}{\Tilde{\mathbf{H}}_{ab}} 
    \\&+ \sum_{ik}\mathbf{W}_{ik}\mathbf{W}_{ia}\frac{\mathbf{\Tilde{H}}_{kb}}{\mathbf{\Tilde{H}}_{ab}}\biggr),
\end{split}
\end{equation}
where $\delta_{ij}$ is the Kronecker delta for indices $i$ and $j$.

Equation (\ref{eq:d2f_dh2}) is a diagonal matrix with strictly positive elements on the diagonal given the 
constraints and is therefore positive semidefinite. From this it follows that 
$F(\mathbf{H}, \Tilde{\mathbf{H}})$ is a convex function, with a global minimum found by 
setting (\ref{eq:df_dh}) equal to 0. Following this procedure and rearranging we
get that $\frac{\partial F(\mathbf{H}, \Tilde{\mathbf{H}})}{\partial \mathbf{H}_{ab}} = 0$
when
\begin{equation}\label{eq:h_nearly_summation}
    \mathbf{H}_{ab} = \mathbf{\Tilde{H}}_{ab} \frac{\sum_{i}[\mathbf{X}_{ib}\mathbf{W}_{ia}]^+}{\sum_{i}[\mathbf{X}_{ib}\mathbf{W}_{ia}]^- + \sum_{ik}\mathbf{W}_{ik}\mathbf{W}_{ia}\mathbf{\Tilde{H}}_{kb}}.
\end{equation}

Setting $\mathbf{\Tilde{H}} = \mathbf{H}$ such that $F(\mathbf{H}, \Tilde{\mathbf{H}}) = J(\mathbf{H})$, 
from  (\ref{eq:auxiliary_def}), this is exactly the $\mathbf{H}$ update rule presented in 
 (\ref{eq:update_nearly}) without weights, thus proving that in the unweighted case 
the Nearly-NMF update rule monotonically decreases (\ref{eq:objective}) with each iteration. 
The weighted case proof follows from using the diagonalized version of the per exposure weights, $[\mathbf{V}_{,j}]^{diag}$, such that the scalar objective in (\ref{eq:objective_nearly_scalar}) becomes
\begin{equation}
\begin{split}
    \chi^2 = &-2\sum_{ijk}[([\mathbf{V}_{,j}]^{diag}\mathbf{X})_{ij}([\mathbf{V}_{,j}]^{diag}\mathbf{W})_{ik}]^+\mathbf{H}_{kj} \\&+ \sum_{ijk}[([\mathbf{V}_{,j}]^{diag}\mathbf{X})_{ij}([\mathbf{V}_{,j}]^{diag}\mathbf{W})_{ik}]^-\mathbf{H}_{kj} \\ &+ \sum_{ijkl}([\mathbf{V}_{,j}]^{diag}\mathbf{W})_{ik}([\mathbf{V}_{,j}]^{diag}\mathbf{W})_{kl}\mathbf{H}_{kj}\mathbf{H}_{ij}
\end{split}
\end{equation}

We can use the same transform as Appendix \ref{proof_shift},
\begin{align} \label{eq:weight_transform}
\begin{split}
    \hat{\mathbf{X}}_{,j} &= [\mathbf{V}_{,j}]^{diag} \mathbf{X}_{,j}, \\
     \hat{\mathbf{W}} &= [\mathbf{V}_{,j}]^{diag} \mathbf{W},
\end{split}
\end{align}
to recover the same form as (\ref{eq:objective_nearly_scalar}). The
only difference is that in this version the value of $\hat{\mathbf{W}}$ will depend
on the index $i$. However, this index will become fixed under the derivative with respect to $\mathbf{H}_{ab}$ 
in (\ref{eq:df_dh}) by 
the delta function $\delta_{ib}$, and the proof proceeds the same way without any further alteration
and eventually recovering a weighted version of  (\ref{eq:h_nearly_summation}), and
therefore a scalar summation form of the $\mathbf{{H}}$ update in (\ref{eq:update_nearly}). This completes
the proof for the $\mathbf{H}$ update rule.

The proof for the $\mathbf{W}$ update rule given in (\ref{eq:update_nearly}) follows the same steps,
both the weighted and unweighted case. Since the outline is the same, it is not necessary to present
the entirety of the steps except to note that the auxiliary function is now
\begin{equation}
\begin{split}
    F(\mathbf{W}, \Tilde{\mathbf{W}}) = &-2\sum_{ijk}[\mathbf{X}_{ij}\mathbf{H}_{kj}]^+\mathbf{W}_{ik}\ \\&+ \sum_{ijk}[\mathbf{X}_{ij}\mathbf{H}_{kj}]^-\frac{\mathbf{W}_{ik}^2 + \mathbf{\Tilde{W}}_{ik}^2}{\mathbf{\Tilde{W}}_{ik}^2}\\ &+ \sum_{ijkl}\mathbf{H}_{kj}\mathbf{W}_{ik}^2\frac{\mathbf{\Tilde{W}}_{il}}{\mathbf{\Tilde{W}}_{ik}}\mathbf{H}_{lj}.
\end{split}
\end{equation}

\section*{Acknowledgments}
The authors would like to thank Dr. D. Kirkby and M. Dowicz for their valuable discussion and input. 
We would also like to thank the reviewers and the associate editor for their valuable comments
in revising the paper.

\bibliographystyle{IEEEtran.bst}
\bibliography{refs.bib}

\begin{thebibliography}{10}
\providecommand{\url}[1]{#1}
\csname url@samestyle\endcsname
\providecommand{\newblock}{\relax}
\providecommand{\bibinfo}[2]{#2}
\providecommand{\BIBentrySTDinterwordspacing}{\spaceskip=0pt\relax}
\providecommand{\BIBentryALTinterwordstretchfactor}{4}
\providecommand{\BIBentryALTinterwordspacing}{\spaceskip=\fontdimen2\font plus
\BIBentryALTinterwordstretchfactor\fontdimen3\font minus \fontdimen4\font\relax}
\providecommand{\BIBforeignlanguage}[2]{{%
\expandafter\ifx\csname l@#1\endcsname\relax
\typeout{** WARNING: IEEEtran.bst: No hyphenation pattern has been}%
\typeout{** loaded for the language `#1'. Using the pattern for}%
\typeout{** the default language instead.}%
\else
\language=\csname l@#1\endcsname
\fi
#2}}
\providecommand{\BIBdecl}{\relax}
\BIBdecl

\bibitem{paatero_positive_1994}
P.~Paatero and U.~Tapper, ``\BIBforeignlanguage{en}{Positive matrix factorization: {A} non-negative factor model with optimal utilization of error estimates of data values},'' \emph{\BIBforeignlanguage{en}{Environmetrics}}, vol.~5, no.~2, pp. 111--126, 1994.

\bibitem{lee_algorithms_2000}
D.~Lee and H.~S. Seung, ``Algorithms for {Non}-negative {Matrix} {Factorization},'' in \emph{Advances in {Neural} {Information} {Processing} {Systems}}, vol.~13.\hskip 1em plus 0.5em minus 0.4em\relax MIT Press, 2000.

\bibitem{pauca_nonnegative_2006}
V.~P. Pauca, J.~Piper, and R.~J. Plemmons, ``\BIBforeignlanguage{en}{Nonnegative matrix factorization for spectral data analysis},'' \emph{\BIBforeignlanguage{en}{Linear Algebra and its Applications}}, vol. 416, no.~1, pp. 29--47, Jul. 2006.

\bibitem{li_non-negative_2013}
Y.~Li and A.~Ngom, ``The non-negative matrix factorization toolbox for biological data mining,'' \emph{Source Code for Biology and Medicine}, vol.~8, no.~1, p.~10, Apr. 2013.

\bibitem{lin_novel_2019}
C.-Y. Lin, L.-W. Kang, T.-Y. Huang, and M.-K. Chang, ``\BIBforeignlanguage{en}{A novel non-negative matrix factorization technique for decomposition of {Chinese} characters with application to secret sharing},'' \emph{\BIBforeignlanguage{en}{EURASIP Journal on Advances in Signal Processing}}, vol. 2019, no.~1, p.~35, Aug. 2019.

\bibitem{blanton_k_2007}
M.~R. Blanton and S.~Roweis, ``\BIBforeignlanguage{en}{\textit{{K}} -{Corrections} and {Filter} {Transformations} in the {Ultraviolet}, {Optical}, and {Near}-{Infrared}},'' \emph{\BIBforeignlanguage{en}{The Astronomical Journal}}, vol. 133, no.~2, pp. 734--754, Feb. 2007.

\bibitem{copar_fast_2019}
A.~Čopar, B.~Zupan, and M.~Zitnik, ``\BIBforeignlanguage{en}{Fast optimization of non-negative matrix tri-factorization},'' \emph{\BIBforeignlanguage{en}{PLOS ONE}}, vol.~14, no.~6, p. e0217994, Jun. 2019, publisher: Public Library of Science.

\bibitem{tsalmantza_data-driven_2012}
P.~Tsalmantza and D.~W. Hogg, ``A {Data}-driven {Model} for {Spectra}: {Finding} {Double} {Redshifts} in the {Sloan} {Digital} {Sky} {Survey},'' \emph{The Astrophysical Journal}, vol. 753, p. 122, Jul. 2012.

\bibitem{zhu_nonnegative_2016}
G.~Zhu, ``\BIBforeignlanguage{en}{Nonnegative {Matrix} {Factorization} ({NMF}) with {Heteroscedastic} {Uncertainties} and {Missing} data},'' \emph{\BIBforeignlanguage{en}{arXiv:1612.06037}}, Dec. 2016.

\bibitem{ren_non-negative_2018}
B.~Ren, L.~Pueyo, G.~B. Zhu, J.~Debes, and G.~Duch{\^e}ne, ``\BIBforeignlanguage{en}{Non-negative {Matrix} {Factorization}: {Robust} {Extraction} of {Extended} {Structures}},'' \emph{\BIBforeignlanguage{en}{The Astrophysical Journal}}, vol. 852, no.~2, p. 104, Jan. 2018, publisher: The American Astronomical Society.

\bibitem{m_nmf-based_2023}
S.~K.~P. M. \emph{et~al.}, ``{NMF-based GPU accelerated coronagraphy pipeline},'' in \emph{Techniques and Instrumentation for Detection of Exoplanets XI}, G.~J. Ruane, Ed., vol. 12680, International Society for Optics and Photonics.\hskip 1em plus 0.5em minus 0.4em\relax SPIE, 2023, p. 1268021.

\bibitem{boulais_unmixing_2021}
A.~Boulais, O.~Berné, G.~Faury, and Y.~Deville, ``\BIBforeignlanguage{en}{Unmixing methods based on nonnegativity and weakly mixed pixels for astronomical hyperspectral datasets},'' \emph{\BIBforeignlanguage{en}{Astronomy \& Astrophysics}}, vol. 647, p. A105, Mar. 2021.

\bibitem{wang_ls-nmf_2006}
G.~Wang, A.~V. Kossenkov, and M.~F. Ochs, ``{LS}-{NMF}: {A} modified non-negative matrix factorization algorithm utilizing uncertainty estimates,'' \emph{BMC Bioinformatics}, vol.~7, no.~1, p. 175, Mar. 2006.

\bibitem{plis_correlated_2009}
S.~M. Plis, V.~K. Potluru, V.~D. Calhoun, and T.~Lane, ``Correlated noise: {How} it breaks {NMF}, and what to do about it,'' in \emph{2009 {IEEE} {International} {Workshop} on {Machine} {Learning} for {Signal} {Processing}}, Sep. 2009, pp. 1--6.

\bibitem{noauthor_sklearndecompositionnmf_nodate}
``\BIBforeignlanguage{en}{sklearn.decomposition.{NMF}},'' \url{https://scikit-learn/stable/modules/generated/sklearn.decomposition.NMF.html} (Accessed: 2024-03-27).

\bibitem{noauthor_nneg_nodate}
``nneg function - {RDocumentation},'' \url{https://www.rdocumentation.org/packages/NMF/versions/0.26/topics/nneg} (Accessed: 2024-03-27).

\bibitem{bolton_sdss_spec1d_2012}
A.~S. {Bolton} \emph{et~al.}, ``{Spectral Classification and Redshift Measurement for the SDSS-III Baryon Oscillation Spectroscopic Survey},'' \emph{The Astronomical Journal}, vol. 144, no.~5, p. 144, Nov. 2012.

\bibitem{brodzeller_qsotemplates_2023}
A.~{Brodzeller} \emph{et~al.}, ``{Performance of the Quasar Spectral Templates for the Dark Energy Spectroscopic Instrument},'' \emph{The Astronomical Journal}, vol. 166, no.~2, p.~66, Aug. 2023.

\bibitem{lyke_dr16q_2020}
B.~W. {Lyke} \emph{et~al.}, ``{The Sloan Digital Sky Survey Quasar Catalog: Sixteenth Data Release},'' \emph{The Astrophysical Journal Supplement Series}, vol. 250, no.~1, p.~8, Sep. 2020.

\bibitem{desi_sv_overview_2023}
{DESI Collaboration} \emph{et~al.}, ``{Validation of the Scientific Program for the Dark Energy Spectroscopic Instrument},'' \emph{arXiv:2306.06307}, Jun. 2023.

\bibitem{guy_desi_pipeline_2023}
J.~{Guy} \emph{et~al.}, ``{The Spectroscopic Data Processing Pipeline for the Dark Energy Spectroscopic Instrument},'' \emph{The Astronomical Journal}, vol. 165, no.~4, p. 144, Apr. 2023.

\bibitem{bailey_empca_2012}
S.~{Bailey}, ``{Principal Component Analysis with Noisy and/or Missing Data},'' \emph{Publications of the Astronomical Society of the Pacific}, vol. 124, pp. 1015--1023, Sep. 2012.

\bibitem{delchambre_wpca_2015}
L.~{Delchambre}, ``{Weighted principal component analysis: a weighted covariance eigendecomposition approach},'' \emph{Monthly Notices of the Royal Astronomical Society}, vol. 446, no.~4, pp. 3545--3555, Feb. 2015.

\bibitem{mcgreer_simqso_2021}
I.~{McGreer}, J.~{Moustakas}, and J.~{Schindler}, ``{simqso: Simulated quasar spectra generator},'' Astrophysics Source Code Library, record ascl:2106.008, Jun. 2021.

\bibitem{dawson_sdss-iv_2016}
K.~S. Dawson \emph{et~al.}, ``\BIBforeignlanguage{en}{The {SDSS-IV} {Extended} {Baryon} {Oscillation} {Spectroscopic} {Survey}: {Overview} and {Early} {Data}},'' \emph{\BIBforeignlanguage{en}{The Astronomical Journal}}, vol. 151, no.~2, p.~44, Feb. 2016.

\bibitem{ren_using_2020}
B.~Ren \emph{et~al.}, ``\BIBforeignlanguage{en}{Using {Data} {Imputation} for {Signal} {Separation} in {High}-contrast {Imaging}},'' \emph{\BIBforeignlanguage{en}{The Astrophysical Journal}}, vol. 892, no.~2, p.~74, Mar. 2020, publisher: The American Astronomical Society.

\bibitem{2020SciPy-NMeth}
P.~Virtanen \emph{et~al.}, ``{{SciPy} 1.0: Fundamental Algorithms for Scientific Computing in Python},'' \emph{Nature Methods}, vol.~17, pp. 261--272, 2020.

\bibitem{mccarty_blind_2020}
R.~J. McCarty, N.~Ronghe, M.~Woo, and T.~M. Alam, ``Blind {Source} {Separation} for {NMR} {Spectra} with {Negative} {Intensity},'' \emph{arXiv:2002.03009}, Feb. 2020.

\bibitem{boashash_chapter_2016}
``Chapter 16 - {Time}-{Frequency} {Methodologies} in {Neurosciences},'' in \emph{Time-{Frequency} {Signal} {Analysis} and {Processing} ({Second} {Edition})}, second edition~ed., B.~Boashash, Ed.\hskip 1em plus 0.5em minus 0.4em\relax Oxford: Academic Press, 2016, pp. 915--966.

\bibitem{kim_subsystem_2003}
P.~M. Kim and B.~Tidor, ``\BIBforeignlanguage{en}{Subsystem {Identification} {Through} {Dimensionality} {Reduction} of {Large}-{Scale} {Gene} {Expression} {Data}},'' \emph{\BIBforeignlanguage{en}{Genome Research}}, vol.~13, no.~7, pp. 1706--1718, Jul. 2003.

\bibitem{harris_array_2020}
C.~R. Harris \emph{et~al.}, ``\BIBforeignlanguage{en}{Array programming with {NumPy}},'' \emph{\BIBforeignlanguage{en}{Nature}}, vol. 585, no. 7825, pp. 357--362, Sep. 2020.

\bibitem{cupy_learningsys2017}
R.~Okuta, Y.~Unno, D.~Nishino, S.~Hido, and C.~Loomis, ``Cupy: A numpy-compatible library for nvidia gpu calculations,'' in \emph{Proceedings of Workshop on Machine Learning Systems (LearningSys) in The Thirty-first Annual Conference on Neural Information Processing Systems (NIPS)}, 2017.

\bibitem{tang_nonnegative_2012}
W.~Tang, Z.~Shi, and Z.~An, ``Nonnegative matrix factorization for hyperspectral unmixing using prior knowledge of spectral signatures,'' \emph{Optical Engineering}, vol.~51, no.~8, p. 087001, Aug. 2012.

\end{thebibliography}

\begin{IEEEbiography}[{\includegraphics[width=1in,height=1.25in,clip,keepaspectratio]{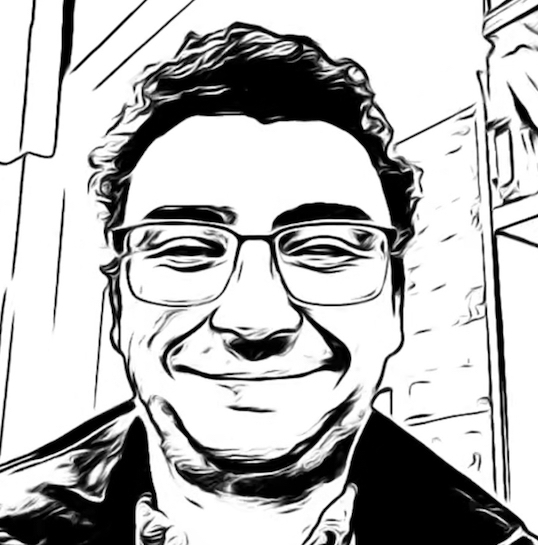}}]{Dylan Green}
received his B.S., and is currently a Ph.D.~candidate, 
in Physics at the University of California, Irvine, in the Department of 
Physics and Astronomy. 
He is a member of the Dark Energy Spectroscopic Instrument, and his research
interests include using physics first principles to improve data processing
for cosmology. \end{IEEEbiography}

\begin{IEEEbiography}[{\includegraphics[width=1in,height=1.25in,clip,keepaspectratio]{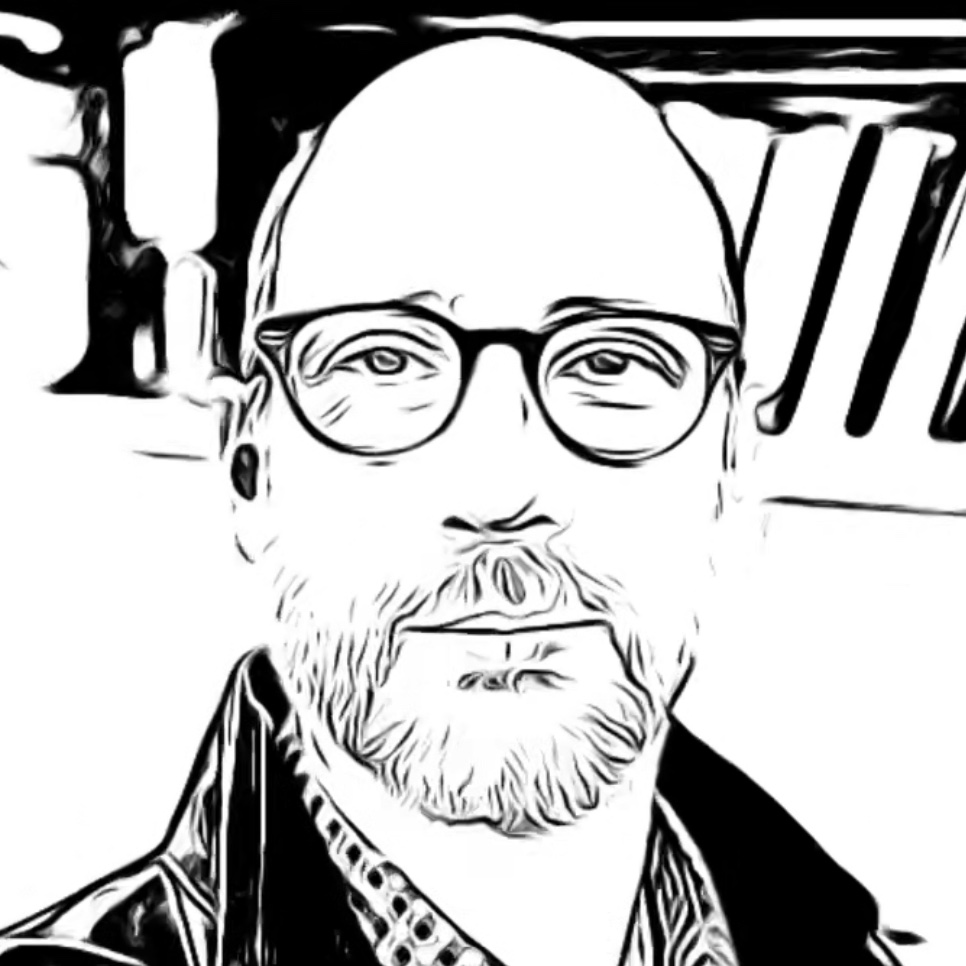}}]{Stephen Bailey}
is a senior software developer in the Physics Division at
Lawrence Berkeley National Lab.
He has been working with large data modeling for over 3 decades,
starting with writing a matchmaking app for his high school.
Along the way he earned a B.S.~degree in Physics from the
University of Washington and a Ph.D.~in Physics from Harvard University.
He currently leads the Data Management team for the Dark Energy Spectroscopic Instrument,
where he enjoys processing raw data into useful data to study cosmology.
\end{IEEEbiography}

\end{document}